\documentclass[useAMS,usenatbib, usegraphicx]{mn2e}

\newcommand{\ebv}{$E(B-V)$}
\newcommand{\Hii}{\mbox{H\,{\sc ii }}}
\newcommand{\Hi}{\mbox{H\,{\sc i }}}
\newcommand{\flux}{\mbox{H$\alpha$ in units of $10^{-15}$ erg s$^{-1}$ cm$^{-2}$ }}
\newcommand{\ageref}{Ages determined by photoionization modeling. A, B, C from \citet{gar97}; F from \citet{gon95} and \citet{gon99}} 
\newcommand{\ebvref}{A,B, C from \citet{gon95}; F from \citet{gon95} (Balmer decrement) and \citet{gon99} (UV continuum)} 
\newcommand{\fnAp}{Ages in Myr from large-aperture broadband photometry for instantaneous starburst (inst) and continuous star formation (cont) models} 
\newcommand{\fnMed}{Median age of \emph{HST}-resolved clusters in the region. These are useful only for regions A, B, D, F; see text for explanation.} 
\newcommand{\units}{FUV, NUV, F380W, F555W, F606W, F814W in units of $10^{-18}$ erg s$^{-1}$ cm$^{-2}$ \AA$^{-1}$} 
\newcommand{\mJy}{$3.6$, $4.5$, $5.8$, $8.0\micron$ in units of mJy} 
\newcommand{\fc}{Fraction of \Hii region luminosity contained in sample clusters in F814W}
\newcommand{\fnLx}{From \citet{smi05a}}
\newcommand{\ccfthirty}{Expected numbers for constant cluster formation over the last 30 Myr at a rate of 1 cluster Myr$^{-1}$}
\newcommand{\ccfhundred}{As above for 100 Myr}

\title[Star Clusters in Arp 284]{Star Clusters in the Interacting Galaxy System Arp 284}
\author[B. W. Peterson et al.]{Bradley W. Peterson$^{1}$\thanks{E-mail:
bwp@iastate.edu (BWP); curt@iastate.edu (CS)}, Curtis Struck$^{1}$, Beverly J. Smith$^{2}$, and Mark Hancock$^{2,3}$\\$^{1}$Department of Physics and Astronomy, Iowa State University, Ames, IA 50011, USA\\
$^{2}$Department of Physics and Astronomy, East Tennessee State University, Johnson City, TN 37614, USA\\
$^{3}$Institute of Geophysics and Planetary Physics, University of California, Riverside, Riverside, CA 92521, USA}
\begin{document}

\date{Accepted YYYY MONTH DAY. Received YYYY MONTH DAY; in original form YYYY MONTH DAY}

\pagerange{\pageref{firstpage}--\pageref{lastpage}} \pubyear{2009}

\maketitle

\label{firstpage}

\begin{abstract}
We present results from a study of proto-globular cluster candidates in the interacting galaxy system Arp 284 (NGC 7714/5) using data from the \emph{Hubble Space Telescope}. Previous studies of the Antennae and M51 have suggested that the majority of young massive star clusters dissolve within 20~Myr due to mass loss. We use the evolutionary synthesis code {\sc starburst99} to estimate ages and extinctions for approximately 175 clusters visible with \emph{HST}. We also use lower-resolution \emph{GALEX} and ground-based H$\alpha$ data to estimate the ages of the giant \Hii regions in which these clusters are found, and compare the \emph{Spitzer} colours of these \Hii regions to those of star forming regions in other interacting systems. The ages are also used to aid in the interpretation of \emph{Chandra} X-ray data. 

Clusters in the tidal tails of NGC 7714 are generally found to have ages less than 20~Myr, though observational limits make the significance of this result uncertain. Older clusters, though not numerous, have nearly the same spatial distribution within the imaged portion of NGC 7714 as young clusters. The cluster population in the bridge connecting the two galaxies appears to be older, but the data in this part of the system is too limited to draw firm conclusions. The ages of the giant \Hii regions in NGC 7714 are generally older than those of their constituent clusters, possibly indicating that the young clusters we detect are surrounded by their dispersed predecessors.
\end{abstract}


\section{Introduction}
The study of star formation in interacting galaxies has been revolutionized by the \emph{Hubble Space Telescope}. One of the early results to emerge from \emph{HST} was the discovery of a class of bright, blue clusters  $\la$ 300 Myr old in the peculiar elliptical galaxy NGC 1275 \citep{hol92}. These young massive clusters (YMCs) were also found in the prototypical merger remnant NGC 7252, with ages ranging from 34--500 Myr, consistent with the age of the merger \citep{whi93}. Many more examples soon followed, in which it was found that the properties of these YMCs implied that they might be young globular clusters, though with higher abundances \citep{con94, hun94}.

In interacting galaxy systems, in which star forming regions are abundant, the high resolution of \emph{HST} has allowed the study of star cluster demographics. Among the results to emerge from such studies is the apparent dissolution of young star clusters, also known as infant mortality. \citet{fall05} used evolutionary synthesis models to estimate the ages of clusters in the Antennae and found that 90 per cent of clusters vanish over each age dex from $10^6$ up to $10^9$ yr, suggesting that  cluster dissolution continues over very long time-scales. In contrast, \citet{bas05a} found in M51 that 70 per cent of clusters dissolve in the first 20~Myr, indicating that disruption is concentrated on short time-scales. Recent spectroscopic observations of star clusters in the Antennae, combined with more careful modeling of the star formation history of the system, does not show the long-term cluster mortality previously claimed \citep{bas09}, thus this issue is still controversial.

Additional evidence for a short dissolution time-scale has been found in comparisons of cluster and field populations. \citet{tre01} obtained UV spectra and evolutionary synthesis models of both clusters and diffuse regions of the dwarf starburst galaxy NGC 5253. Unlike the clusters, the diffuse region spectra lacked indications of O-type stars. It was suggested that stars form entirely within clusters, which dissolve on $\sim10$~Myr time-scales, preventing short-lived O stars from ever entering the field population.

Early-stage disruption has been ascribed largely to the loss of gas left over from star formation, which is swept out of the cluster by stellar winds and massive star supernovae. The resulting mass loss drops the escape velocity below the stellar velocities, causing the cluster to expand and decrease in surface brightness. The departing stars disperse into the field population in $\sim$few $10^{7}$ yr \citep{bas06,goo06}. On longer time-scales ($\sim10^{8}$~yr and longer), interactions with giant molecular clouds \citep{gie06} and tidal effects are expected to become significant disruption mechanisms.

If YMCs are progenitors of globular clusters, some of them must survive their most massive stars, as well as external effects that may lead to dissolution \citep{deg07}. Interacting galaxy systems have high numbers of YMCs, making them ideal targets for cluster population studies. To date, too few interacting systems have been observed to definitively determine how rapidly and by what mechanism massive clusters dissolve. Further, both the Antennae and M51 are advanced interactions with complex star formation histories. It is necessary to confirm infant mortality in early-stage interactions where the star formation history is simpler, with a particular emphasis on covering a wide range of tidal environments.

In this paper we focus on the early-stage interacting system Arp 284, which at a distance of  $38.6\pm2.7$~Mpc ($1\arcsec = 190$ pc) according to the NASA/IPAC Extragalactic Database (NED) has nearly 175 detected star forming regions in archival \emph{HST} images. We use \emph{HST} colours and the evolutionary synthesis code {\sc starburst99} ({\sc sb99}) to obtain age and reddening estimates for star forming regions. We also use lower-resolution UV data from the \emph{Galaxy Evolution Explorer} (\emph{GALEX}; \citealt{mar05}) and ground-based H$\alpha$ to obtain age estimates for larger \Hii regions, and mid-IR colours from the \emph{Spitzer Space Telescope} Infrared Array Camera (IRAC; \citealt{wer04,faz04}) to compare the \Hii regions to IR emission clumps found in other interacting systems. These age results are used to aid in the interpretation of the X-ray data of \citet{smi05a},which was obtained using \emph{Chandra} \citep{wei02}.

\section[]{Tidal Star Forming Environments of Arp 284}
\begin{figure}
\includegraphics[width=84mm]{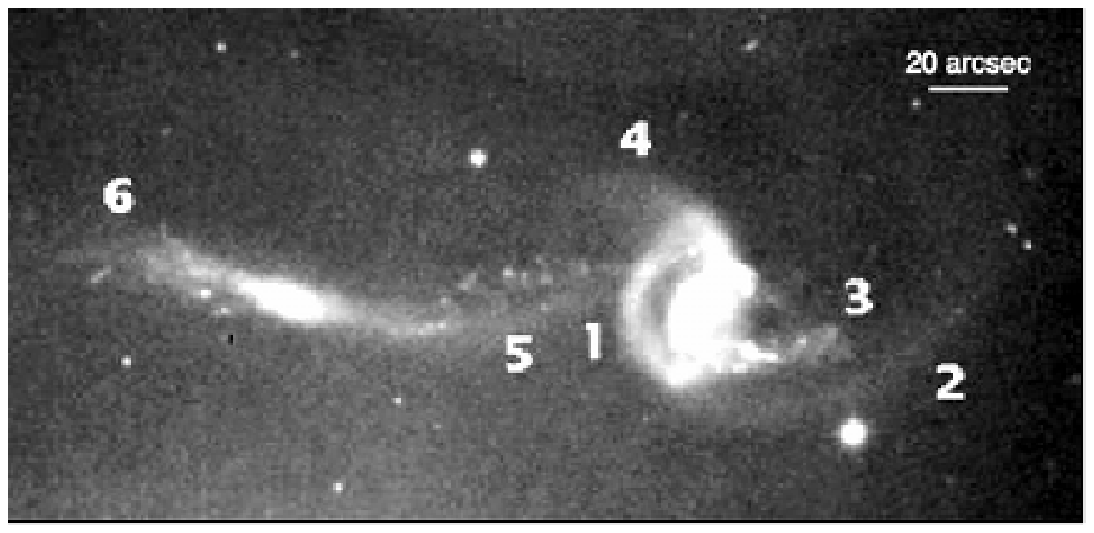}
\caption{Optical photograph of Arp 284 from \citet{arp66}, with numbers labeling the prominent tidal features. North is up and east to the left. The field of view is $4.5\arcmin\times2.2\arcmin$.
\label{fig:arp}}
\end{figure}

\begin{figure}
\includegraphics[width=84mm]{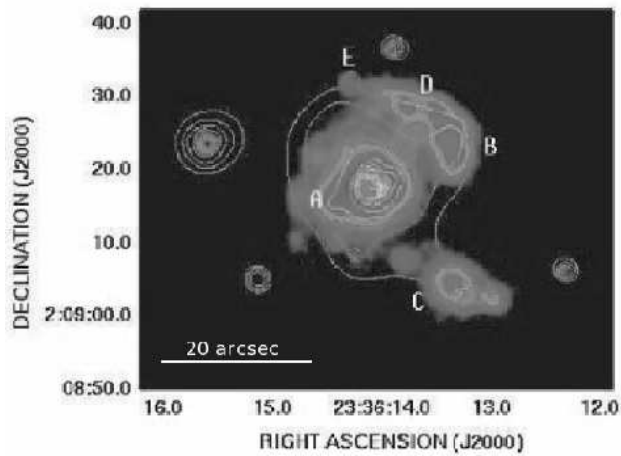}
\caption{NGC 7714 disc indicating \Hii regions from \citet{smi05a}. The colour scale shows H$\alpha$. The contours are a Chandra 0.3--8 keV x-ray map. 
\label{fig:chandra}}
\end{figure}

The peculiar morphology of the Arp 284 system results from an off-centre collision between two disc galaxies, with the primary (NGC 7714) having roughly three times the mass of the companion (NGC 7715). The time of closest approach was $\sim100$~Myr ago. For a discussion of the interaction history of the system, the reader is directed to the models of \citet{str03}. In this section we summarize previous observations of the Arp 284 system that relate to its star formation history. Most of the tidal features are labelled in Fig.~\ref{fig:arp}, while the \Hii region complexes in the primary disc are labelled in Fig.~\ref{fig:chandra}.

\subsection{Nucleus of NGC 7714}
The NGC 7714 nucleus was the first object to be described using the term `starburst,' although there are earlier references in the literature to `bursts of star formation'  \citep{b21}. Strong H$\alpha$  emission indicates a large nuclear \Hii region and ongoing star formation \citep{gon95,smi97}. O and B stars are required to fit the UV spectrum, with a best fitting age $\sim$5~Myr and a burst of star formation favored over continuous star formation \citep{gon99}.

\citet{lan01} found that a second, older nuclear population is suggested by strong absorption bands of CO in the near-IR. In particular, the \emph{K}-band spectrum is compatible with a combination of red supergiants, AGB stars, and RGB stars. They concluded that a two-component model is probably an oversimplification of the nuclear star formation history, preferring a star formation rate that has decreased over several $10^{8}$ yr, either through a smooth exponential decrease or a series of successively smaller bursts. The burst model best fits the observed supernova rate with a burst 15--50~Myr ago, in addition to the most recent burst 5~Myr ago.

\emph{Infrared Space Observatory} (\emph{ISO}) and \emph{Spitzer} mid-IR spectra of the nuclear region show strong features from polycyclic aromatic hydrocarbons (PAHs) as well as bright forbidden lines \citep{oha00, bra04}. The metallicity of NGC 7714 peaks in the nucleus, with a value near $0.5Z_{\sun}$, and generally decreases with distance from the nucleus \citep{gon95}.

\subsection{Stellar Ring of NGC 7714 (Feature~1)}
NGC 7714 has a partial ring-like structure to the east. This feature does not show prominent H$\alpha$ emission, reflecting a lack of ongoing star formation, while the red continuum suggests an old stellar population \citep{gon95,smi97}. Low levels of \Hi emission indicate that there is little hydrogen gas \citep{smi97}.

\subsection{Southern Tails of NGC 7714 \\* (Features~2~and~3)}
\begin{figure}
\includegraphics[width=84mm]{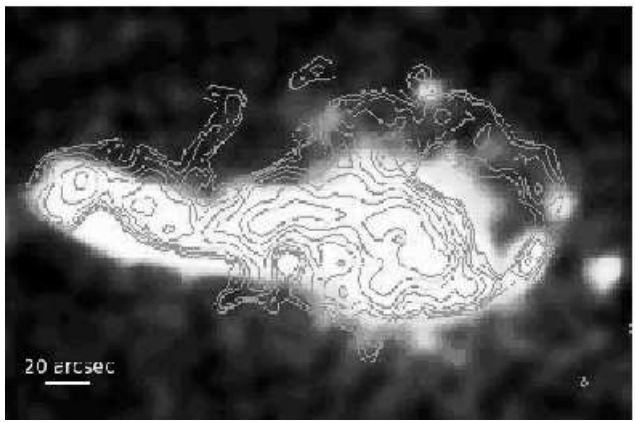}
\caption{Smoothed \emph{GALEX} FUV (colour) overlaid with VLA \Hi (contours). Note that the western tail loops back to the main galaxy, and this ring-like feature is present in both \Hi and UV. The field of view is $5.1\arcmin\times3.5\arcmin$.
\label{fig:loop}}
\end{figure}

Two parallel  tidal tails lie to the southwest of the NGC 7714 nucleus. The outer southern tail, faintly visible in optical images, is revealed by 21 cm maps to be part of a large \Hi loop which reconnects with the galaxy in the northern tail \citep{b23,smi97}. The loop is also visible in \emph{GALEX} FUV, which is shown smoothed with overlaid \Hi contours in Fig.~\ref{fig:loop}. This gas loop is an entirely separate structure from the inner southern tail, which hosts significant ongoing star formation.

The base of the inner southern tail just south of the nucleus contains \Hii region A \citep{smi05a}. This region is visible in 20 cm radio continuum  \citep{gon95,smi97}. The \emph{HST} images reveal numerous bright clusters along this structure, with another large \Hii region at the edge of the field (region C; \citealt{gon95}), coincident with a bright X-ray source \citep{smi05a}. The ages of regions A and C have been estimated by \citet{gar97} at $5.0\pm0.5$ and $4.5\pm0.5$~Myr based on photoionization models. These regions also appear in \emph{Spitzer} images \citep{smi07}, to be discussed below.

\citet{gon95} found H$\alpha$ region A to have a metallicity comparable to the nucleus, about $0.5Z_{\sun}$. Region C, at a larger distance from the nucleus, shows a lower metallicity of $0.25Z_{\sun}$.

\subsection{Northern Tail of NGC 7714 (Feature~4)}
The northern tail of the primary galaxy is visible but not particularly prominent in optical photographs. Near the base of the tail, to the northwest of the nucleus, are H$\alpha$ regions B, D, and E. These regions are also detected in 21 cm \Hi emission and red continuum \citep{smi97}. \emph{HST} resolves these regions into several smaller groups of clusters. The age of region D has been estimated at $3.5\pm0.5$~Myr using photoionisation modeling \citep{gar97}.

H$\alpha$ region B has been found to have lower metallicty than the nucleus, about $0.25Z_{\sun}$ \citep{gon95}.

\subsection{Bridge (Feature~5)}
The bridge consists of a northern and a southern component. The northern bridge shows `clumpy' emission from \Hi gas, and several H$\alpha$ peaks are also found, offset slightly to the south, indicating ongoing star formation. The southern component lacks \Hi gas and H$\alpha$ emission, but is readily seen in the red continuum, likely indicating an older stellar population  \citep{smi97}. No metallicity data is currently available for the bridge.

\subsection{Nucleus of NGC 7715}
The lack of significant H$\alpha$ emission in the NGC 7715 nucleus indicates that there is no ongoing massive star formation \citep{gon95,smi97}. Emission from the 21 cm \Hi radio line is also minimal, and offset to the north of the optical galaxy. The nucleus is comparatively bright in the red continuum, suggesting an aged stellar population \citep{smi97}. Spectral studies of the nucleus show a post-starburst spectrum, again confirming the lack of ongoing star formation \citep{ber93}.

\subsection{Eastern Tail of NGC 7715 \\* (Feature~6)}
The eastern tail of NGC 7715 is faintly seen in optical images. Radio maps show an \Hi counterpart, offset slightly to the north \citep{b23,smi97}. Massive star formation is thought to be weak based on the lack of H$\alpha$ emission. This feature has not been studied in detail, and it falls outside the field of the \emph{HST} images. This poor coverage is unfortunate, because the models suggest that this tail curves behind the NGC 7715 disc, passes behind the bridge, and comes back into view as the faint outer southern tail of NGC 7714 \citep{str03}. In this scenario, the tail contains a significant fraction of the original mass from the NGC 7715 disc \citep{str03}.

\section{Observations}
\subsection{\emph{HST} observations}
The \emph{HST} Wide Field and Planetary Camera 2 (WFPC2) has observed the Arp 284 system through several observing programs between 1995 May 13 and 2001 July 10. These programs targeted either the starburst nucleus or SN 1999dn in the disc of NGC 7714. The disparate nature of the observations means that many of the bright clusters are not visible in all filters, and in some cases the clusters were not found on the same chip across all filters. The field of view is $36\arcsec\times36\arcsec$ on the PC chip and $80\arcsec\times80\arcsec$ for the WF chips, with pixel sizes of $0.05\arcsec$ pixel$^{-1}$ on the PC chip and $0.1\arcsec$ pixel$^{-1}$ on the WF chips. The images and exposure times are summarized in Table~\ref{tab:hst}. Note that the number in the filter name approximately corresponds to the central wavelength of the filter in nanometers. The available filters were F380W ($\sim U$), F555W ($\sim V$), F606W ($\sim R$), and F814W ($\sim I$).

The NGC 7714 nucleus and most of the inner southern tail are visible on the PC chip in all filters. One image in both F380W and F814W had to be discarded due to excessive cosmic ray contamination. This left three good images in the F380W filter, two in F555W, one in F606W, and three in F814W. In F814W, \Hii region C was visible in two images on the WF4 chip. 

The entire northern tail was visible on the PC chip in the F380W and F606W filters. In F555W and F814W, only 12 clusters were visible on the PC chip, with the remainder visible on WF4 in F814W.

The bridge was on the seam between the WF3 and WF4 chips in all four F380W images, so some of the clusters could not be measured in this band. In the other filters, it was always on the WF3 chip, with two images in F555W, one in F606W, and two in F814W.

\subsection{\Hii region observations}
\begin{figure}
\includegraphics[width=84mm]{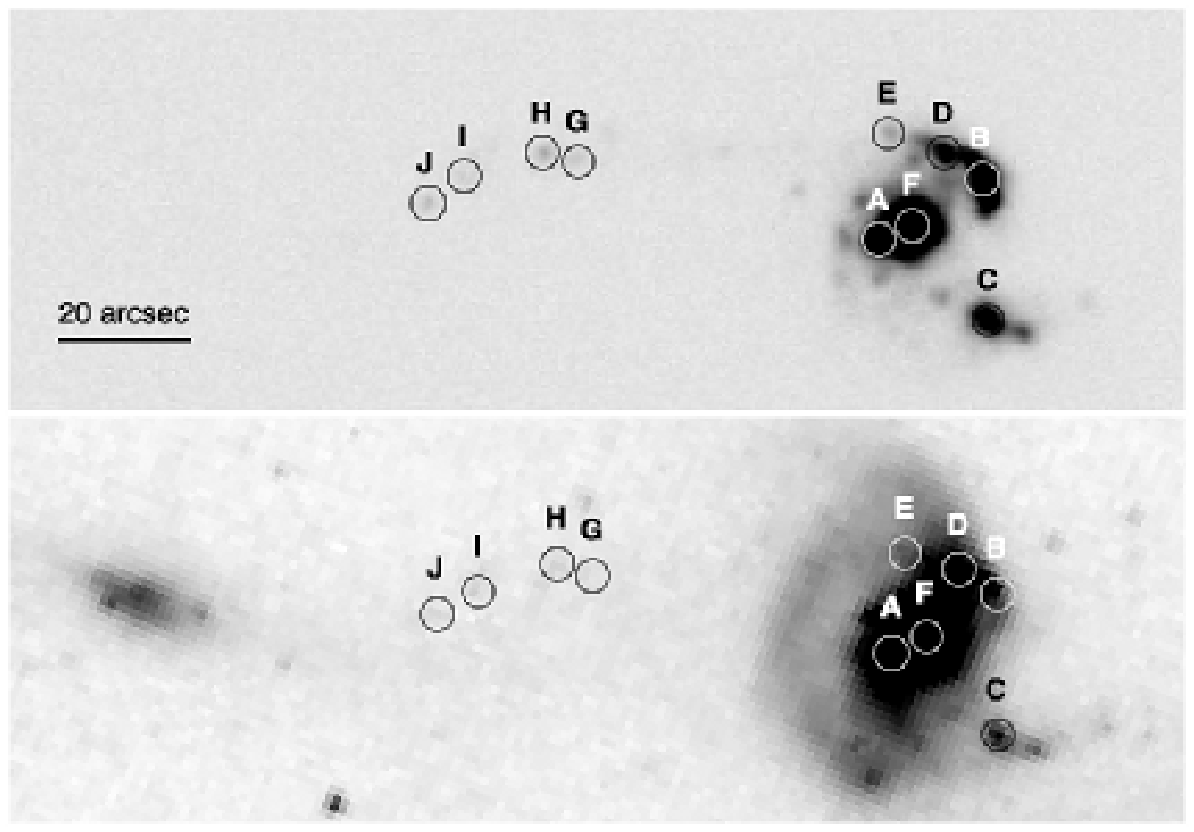}
\caption{\Hii regions in H$\alpha$ (top) and \emph{Spitzer} $3.6\micron$ (bottom). The aperture radii have an angular size of  $2.45\arcsec$, and are also shown on the \emph{HST} images in Figs.~\ref{fig:regions}--\ref{fig:bridge}. The field of view is $2.4\arcmin\times0.9\arcmin$.
\label{fig:spitz}}
\end{figure}

Archival \emph{GALEX} and \emph{Spitzer} IRAC data are available for the Arp 284 system. The \emph{GALEX} images increase the number of colours that can be fit with {\sc sb99} models, while \emph{Spitzer} colours can be compared to those of star forming regions in other interacting systems. These instruments have significantly lower resolution than \emph{HST}, but still allow the \Hii regions A--E to be resolved, as shown in Fig.~\ref{fig:spitz}. Each of these regions contains numerous star clusters resolved by \emph{HST}. For clarity, the point-like sources in the \emph{HST} data will be called `clusters' while these larger structures will be called `\Hii regions.' With H$\alpha$ luminosities of $10^{39}-10^{41}$ erg s$^{-1}$ \citep{smi05a}, these are properly classified as `giant \Hii regions' \citep[e.g.][]{ken84}.

The \Hii regions were located in an H$\alpha$ map, which was then registered to the other images. The H$\alpha$ and accompanying \emph{R}-band images were obtained using the 1.8 m Perkins Telescope of the Ohio State University, using a CCD with a pixel size of $0.49\arcsec$ pixel$^{-1}$. These data were published previously in \cite{smi97}.

\emph{GALEX} observations are available in both the FUV and NUV bands, with effective wavelengths of 1516 and 2267 \AA . The pixel size is $1.5\arcsec$ pixel$^{-1}$, which makes even these relatively large \Hii regions difficult to identify. The integration times are 4736 s in FUV and 520 s in NUV.

IR data from \emph{Spitzer} are available for both the IRAC and Multiband Imaging Photometer for Spitzer (MIPS) instruments. The resolution in the MIPS bands is far too low for cluster studies, and there are pronounced artifacts due to the point spread function because of the brightness of the nucleus, so only the broadband IRAC bands (3.6, 4.5, 5.8, and 8.0$\micron$) were used. The data reduction and mosaicing for the \emph{Spitzer} data are described in \citet{smi07}.

\section{\emph{HST} Data Analysis}
\subsection{Data Reduction}
All of the images were processed by the standard Hubble pipeline. The small number of exposures available in some data sets made cosmic ray removal a non-trivial task, since median combining two images is ineffective. First, we used the {\sc iraf}\footnote{{\sc iraf}  is distributed by the National Optical Astronomy Observatories, which are operated by the Association of Universities for Research in Astronomy, Inc., under cooperative agreement with the National Science Foundation.} task \emph{cosmicrays} to remove events a few pixels in size that were well above background. Next, images were blinked to identify remaining cosmic rays, which were removed manually using the \emph{fixpix} task. Finally, images were combined using the \emph{imcombine} task. For F606W, only one image was available, so other filters were used for comparison.

\subsection{Cluster Selection}
Clusters were selected using the {\sc iraf} task \emph{daofind}. The background level $\sigma$ was determined using \emph{imstat} in a relatively dark part of the image. The nuclear region is  much brighter than the rest of the disc, so it was necessary to use a higher background for cluster detection, measured at the outer edge of the nucleus. We used \emph{daofind} in each filter and then merged the lists to obtain the final sample. Clusters detected in only two filters were not included in the sample.

We detected 16 clusters in the nucleus at the $5\sigma$ level. The nucleus is quite crowded, so no attempt was made to push the detection threshold any lower. The lowest detected flux was $2.28\times10^{-17}$ erg s$^{-1}$ cm$^{-2}$ \AA$^{-1}$ in F814W.

In the southern tail, 60 clusters were detected at the $5\sigma$ level and another 34 were detected at the $3\sigma$ level. These levels correspond to fluxes of 2.17 and $1.32\times10^{-19}$ erg s$^{-1}$ cm$^{-2}$ \AA$^{-1}$, respectively, in F814W. Before using the software, clusters in this part of the system were selected by eye. The $3\sigma$ set matched well with the results of the visual inspection, and were retained for the analysis.

In the northern tail, \emph{daofind} detected 60 clusters at the $5\sigma$ level, with no additional clusters detected at the $3\sigma$ level. This is probably due to crowding in the major complexes, which resulted in multiple detections on single clusters at lower detection thresholds. These clusters may in fact represent multiple unresolved star clusters. In a few cases, multiple resolved clusters had to be measured in a single aperture because they were too close together to measure without aperture overlap.

Cluster selection was difficult in the bridge due to the higher background level in this part of the system and the lower resolution of the WF3 chip compared to the PC chip. For these reasons, it was necessary to lower the detection threshold in the bridge to $2\sigma$, allowing the detection of 34 clusters. The faintest flux detected in F814W was $2.17\times10^{-19}$ erg s$^{-1}$ cm$^{-2}$ \AA$^{-1}$. Note that this is higher than the $3\sigma$ level on the PC chip. Only 23 of the clusters could be measured in F380W.

\subsection{Photometry of clusters}
Photometry was performed using the {\sc iraf} task \emph{phot} in the {\sc daophot} package, with magnitudes calibrated to the Vega magnitude system using the zero point values given in the WFPC2 Data Handbook. The magnitudes and ages for the cluster sample are given in Table~\ref{tab:clage}, while the extinctions and mass estimates are given in Table~\ref{tab:clebv}. The determination of the ages and extinctions will be discussed in \S~\ref{sec:aged}. The mass determination will be discussed in \S~\ref{sec:mass}.

The tidal tails generally fell on the PC chip. For these clusters, we used aperture radii of 3 or 4 pixels for single clusters, depending on crowding, and 6 pixels in six cases when barely-resolved clusters had to be measured in a single aperture. Sky brightness for background subtraction was determined by the mode of the pixel values in an annulus of 10--15 pixels, centered on the cluster. In F814W, most of the northern tail was on the WF4 chip. These were measured using apertures with 2-pixel radii, or 3-pixel apertures for the poorly resolved groups, with sky subtraction determined by the mode in a 5--8 pixel annulus. These sizes were chosen because the WF chips have approximately twice the angular size of the PC chip, so the same area of sky was measured regardless of which chip a cluster fell on. Very little of the northern tail was visible in F555W. 

The nucleus was always on the PC chip, but required the use of smaller apertures with radii of 2 or 3 pixels due to crowding. Background levels were determined by the mode in a 10--15 pixel annulus.

The bridge was generally on the WF3 chip. In F380W, the bridge unfortunately straddled the WF3 and WF4 chips, so a few clusters could not be observed in this filter. The aperture radius was 3 pixels, with background determined by the median in an 8--13 pixel annulus. This aperture size was selected because bridge clusters are often isolated, but there are several areas where multiple emission sources are packed closely together. In such cases, \emph{daofind} generally detected only one or two of the brightest sources, which were always sufficiently separated to be measured individually.

Aperture corrections could not be determined on the images. It was therefore necessary to use the standard corrections prescribed by \citet{hol95}. The corrections for WF3 were used for all of the wide field chips. Standard corrections are not given for the F380W or F606W filters, so corrections for the F336W and F555W, respectively, were used. The corrections were typically $\la~0.3$~mag.

\subsection{\Hii region photometry} 
In addition to the five regions A--E, we measured the nucleus (designated region F) and four regions (G--I) in the northern part of the bridge, shown in H$\alpha$ and \emph{Spitzer} $3.6\micron$ in Fig.~\ref{fig:spitz}. The \emph{GALEX} and \emph{Spitzer} images were registered to the \emph{HST} data so that age estimates could be obtained using the total flux in the \emph{HST} bands. This also allowed identification of the clusters making up the \Hii regions. Most of region C was off frame in all of the \emph{HST} bands except F814W. Region G could not be measured at $4.5\micron$ because of a cosmic ray.

The aperture radius was selected as $2.45\arcsec$ (5 pixels on the H$\alpha$ image) in all instruments to ensure that the same area of sky was represented at all wavelengths. The sky annulus widths were $1.82\arcsec$ in \emph{HST}, $2.45\arcsec$ in H$\alpha$, $4.5\arcsec$ in \emph{GALEX}, and $3.6\arcsec$ in \emph{Spitzer}. The measurements were not aperture-corrected because the \emph{HST} apertures were largely empty, and the emission regions were highly pixelated in \emph{GALEX} and \emph{Spitzer}.

Table~\ref{tab:hiiphot} summarizes the photometric measurements. The H$\alpha$ and \emph{R} bands were calibrated to standard units using the results of \citet{jam04}. The ratio of the H$\alpha$ flux to \emph{R} flux density provided an estimate of the equivalent width of the H$\alpha$ line, EW(H$\alpha$). This equivalent width is included in the {\sc sb99} models and provides another method for determining the ages of the emission regions. Ages are presented in Table~5 and extinctions in Table~\ref{tab:hiiebv}. These will be discussed in \S~\ref{sec:hiiage}.

\section{\emph{HST} Results}
\subsection{Cluster colours}
\begin{figure}
\includegraphics[width=173mm]{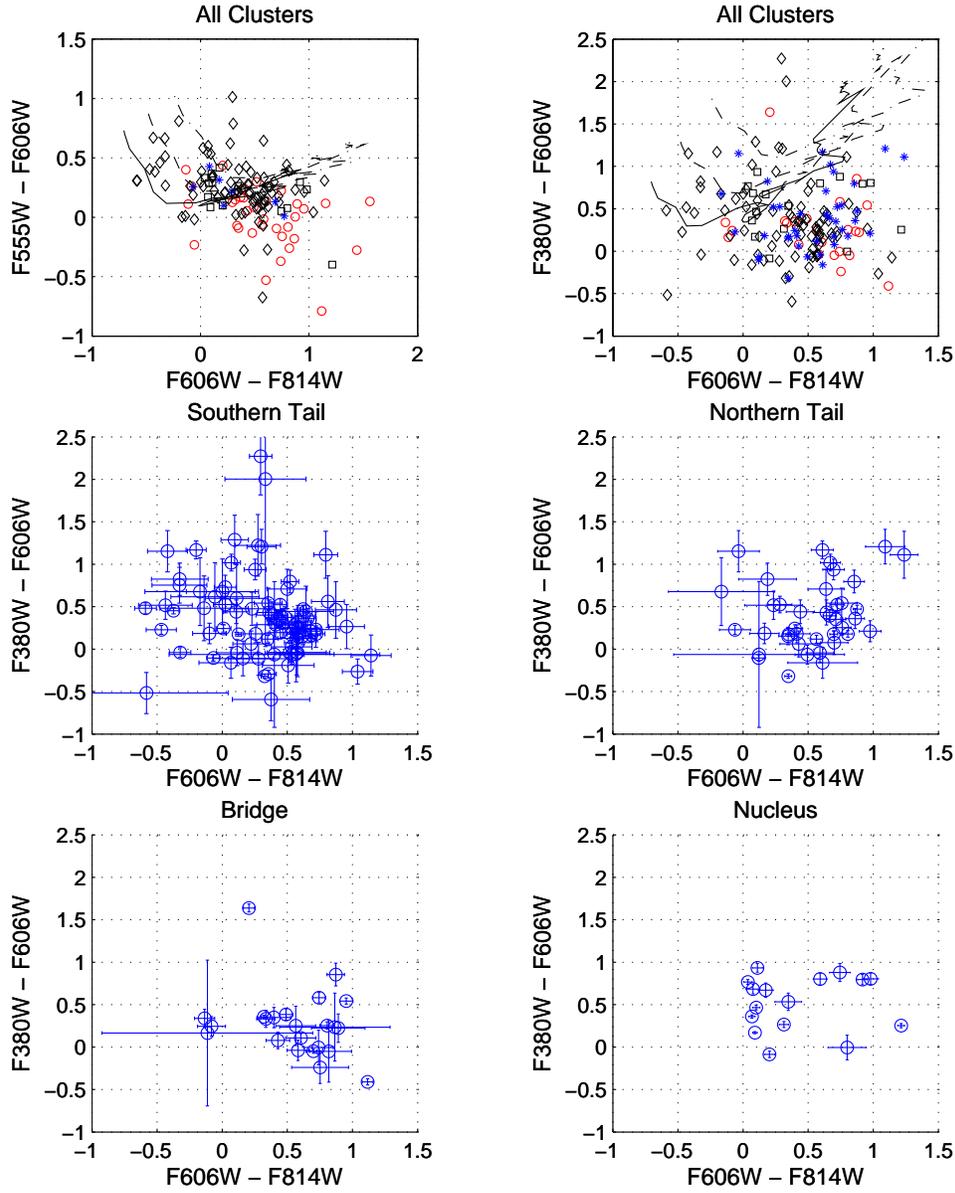}
\caption{In the top panels, red circles represent clusters from the bridge, black diamonds the southern tail, blue asterisks the northern tail, and black squares the nucleus. The curves show {\sc sb99} instantaneous star formation models with $0.2Z_{\sun}$ metallicity for $E(B-V)=0$ (solid curve), 0.24 (dashed curve), and 0.5 (dot-dashed curve). These curves give a sense of the effect of reddening, but the metallicity is not representative of the nucleus or parts of the southern tail. The middle and bottom four panels show the same colours as the upper right panel separated by region, with error bars.
\label{fig:cc}}
\end{figure}

Colour-colour diagrams are shown in Fig.~\ref{fig:cc}. In the top left panel, we plot the magnitude in the F555W filter minus the magnitude in the F606W filter (F555W$-$F606W) versus (F606W$-$F814W) for all detected clusters for which these colours were available. The curves show {\sc sb99} instantaneous star formation models with $0.2Z_{\sun}$ metallicity for $E(B-V)=0$ (solid curve), 0.24 (dashed curve), and 0.5 (dot-dashed curve) for ages from $10^6$--$10^{10}$ yr. The youngest ages are at the lowest values of (F606W$-$F814W). The models are shown to give a sense of the effect of reddening, and it should be noted that $0.2Z_{\sun}$ is not the observed metallicity for the nucleus and parts of the southern tail. The bridge population is generally bluer in (F555W$-$F606W) and redder in (F606W$-$F814W) than the populations of the other regions. The southern tail has the opposite tendency, leaning red in (F555W$-$F606W) and blue in (F606W$-$F814W). Only a few clusters in the northern tail were on frame in F555W; those that were tend toward more intermediate values in both colours. The nuclear population is similar.

The top right panel shows all detected clusters in (F380W$-$F606W) versus (F606W$-$F814W), along with {\sc sb99} models. The youngest ages on the curves appear at the lowest values of (F606W$-$F814W). The middle and bottom panels show the same colours, with the different regions plotted separately and error bars shown. The differences between the populations are less striking in this colour combination than the one discussed above, and arise primarily in (F606W$-$F814W). 

The median statistical uncertainties in the northern tail are 0.081 and 0.095 mag for (F380W$-$F606W) and (F606W$-$F814W), respectively. In the southern tail, the median uncertainties are 0.153 and 0.107 mag; in the nucleus, 0.050 and 0.040 mag; and in the bridge, 0.167 and 0.123 mag. The uncertainties are large in the southern tail in part because it includes clumps detected at the $3\sigma$ level. If only $5\sigma$ level clumps are included, the median uncertainties are 0.114 in (F380W$-$F606W) and 0.088 in (F606W$-$F814W).

\subsection{Age determination} \label{sec:aged}
Age and reddening estimates were obtained using the {\sc starburst99} v5.1 evolutionary synthesis code \citep{lei99, vaz05}. We modeled the colours of the clusters assuming a single, instantaneous burst of star formation with a \citet{kro02} initial mass function over the range 0.1--$100M_{\sun}$. Spectral energy distributions (SEDs) were calculated for cluster ages of 1~Myr to 10~Gyr and included contributions from H$\alpha$ emission. The SEDs were reddened from 0--2 mag of $E(B-V)$ using the reddening law of \citet{cal94}. These model spectra were convolved with the bandpasses of the four \emph{HST} filters. The model age and extinction providing the best fit to the observations was determined using $\chi^2$ minimization

Uncertainties in both the ages and extinctions were determined as in \citet{smi08}, and are driven largely by the degeneracy between age and reddening. Breaking this degeneracy is particularly difficult for clusters with coverage in only three filters. Such clusters generally have larger age uncertainties than those with that were covered in all four filters.
	
Models were run for three different metallicities: $Z_{\sun}$, $0.4Z_{\sun}$, and $0.2Z_{\sun}$. The $0.4Z_{\sun}$ model is representative of the spectroscopically measured abundances in the nucleus and \Hii region A, while the $0.2Z_{\sun}$ model is close to that of \Hii regions B and C. The solar composition models were used primarily for comparison.

Several colour combinations were used because not all clusters were available in all colours. Age estimates obtained from {\sc sb99} have been found to be most reliable when (1) large numbers of colours are used and (2) the bluest available combination of colours are used. \emph{U} coverage is particularly important \citep[e.g.][]{han08}. We adhered to these results in determining which of the available colour combinations would provide the best age estimate for each cluster at each metallicity.

\begin{figure}
\includegraphics[width=150mm]{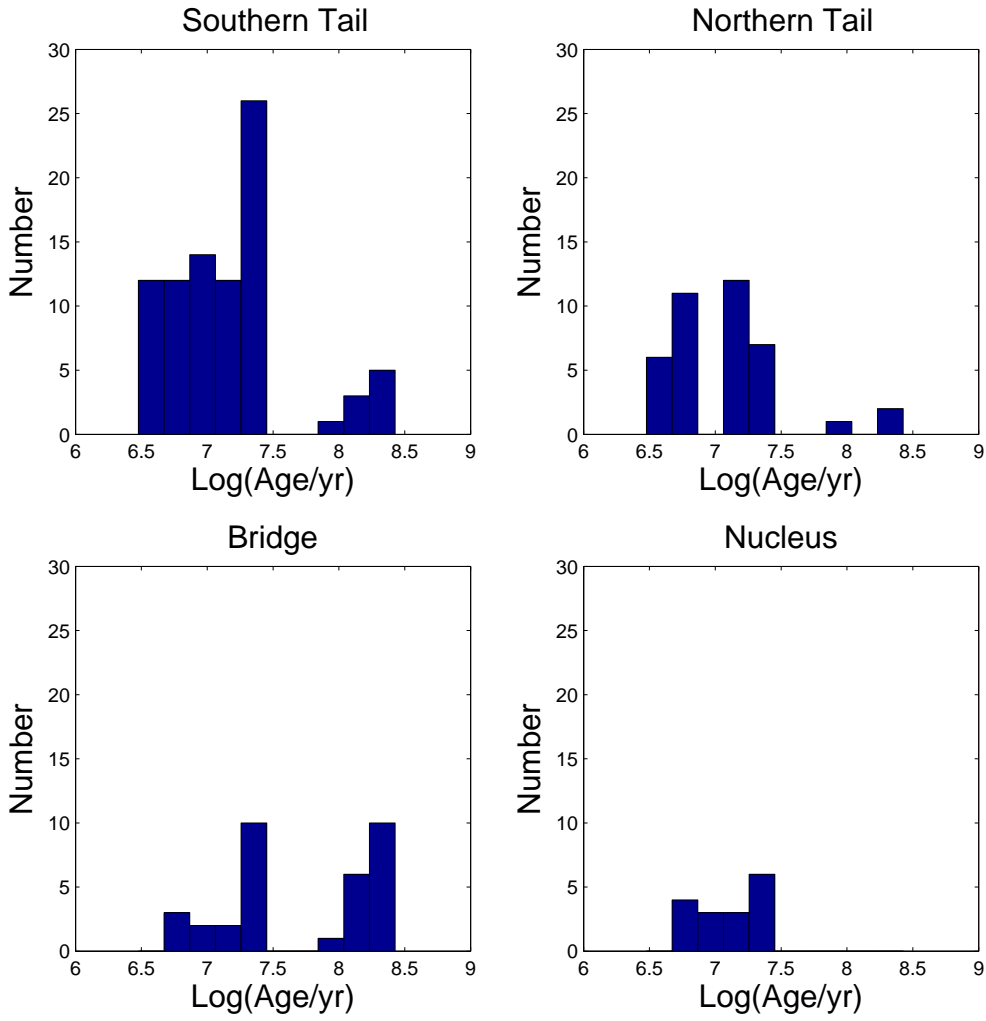}
\caption{Histograms of cluster ages from $0.2Z_{\sun}$ models in the southern tail (top left panel), northern tail (top right panel), and bridge (bottom left panel), and from the $0.4Z_{\sun}$ model in the nucleus (bottom right panel).
\label{fig:ageHist}}
\end{figure}

\begin{figure}
\includegraphics[width=84mm]{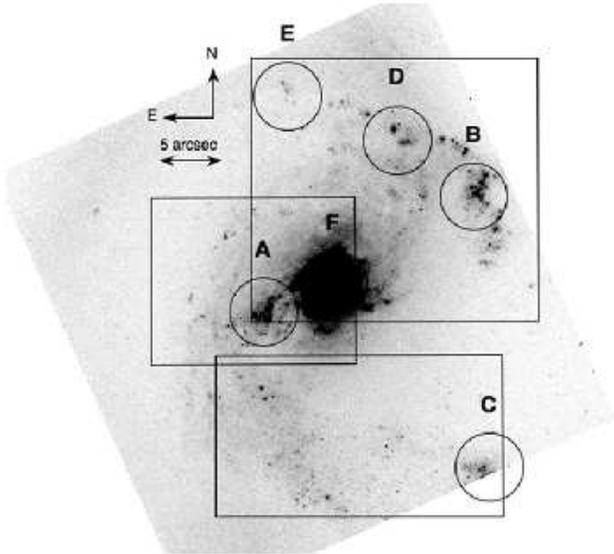}
\caption{F606W image of NGC 7714 with northern, central, and southern areas indicated. These areas are examined more closely in Figs.~\ref{fig:north}--\ref{fig:bridge}. Several \Hii regions identified in lower-resolution data are also indicated. Note that region C is partially off frame toward the bottom right of the box indicating the southern area.
\label{fig:regions}}
\end{figure}

\begin{figure}
\includegraphics[width=84mm]{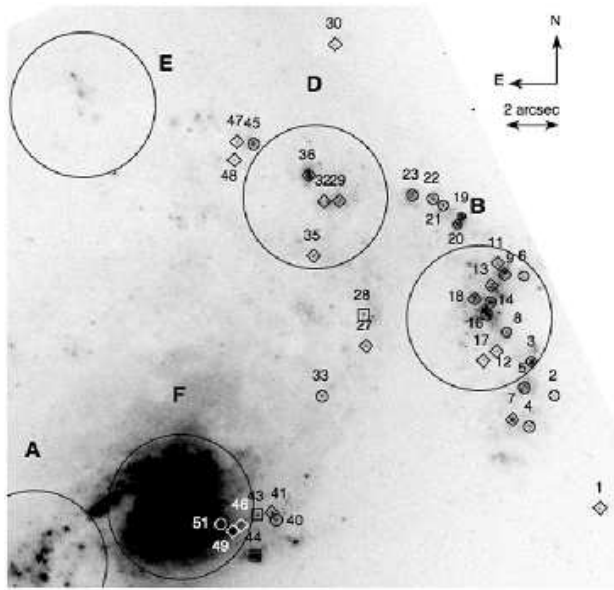}
\caption{F606W image of the northern tail of NGC 7714, with ages indicated for all detected clusters for the $0.2Z_{\sun}$ models. Circles represent ages 1--10 Myr, diamonds 11--22 Myr, squares $\geq 23$ Myr. The metallicities of \Hii regions B and D have been measured at $0.25Z_{\sun}$, while the nucleus has metallicity $0.5Z_{\sun}$. All clusters shown are considered part of the northern tail.
\label{fig:north}}
\end{figure}

\begin{figure}
\includegraphics[width=84mm]{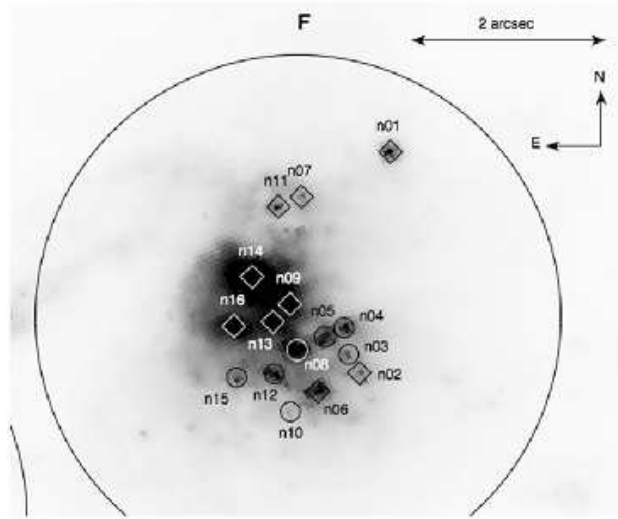}
\caption{F606W image of the NGC 7714 nucleus with ages from the $0.4Z_{\sun}$ model indicated for all detected clusters. Circles represent ages 1--10 Myr, diamonds 11--22 Myr, squares $\geq 23$ Myr. The metallicity has been measured at $0.5Z_{\sun}$.
\label{fig:nuc}}
\end{figure}

\begin{figure}
\includegraphics[width=84mm]{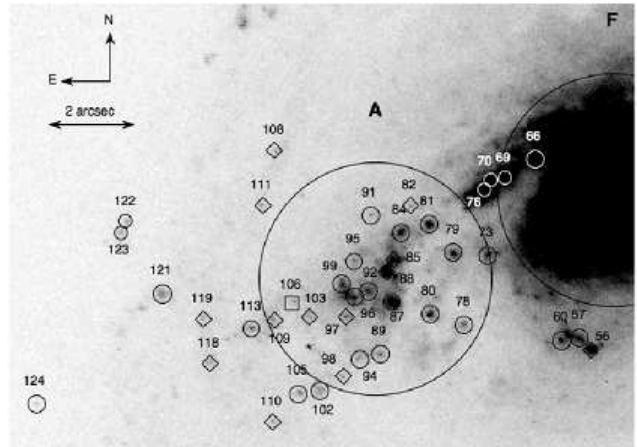}
\caption{F606W image of the central region of NGC 7714 with ages from the $0.4Z_{\sun}$ model indicated for all detected clusters. Circles represent ages 1--10 Myr, diamonds 11--22 Myr, squares $\geq 23$ Myr. The nucleus has a measured metallicity of $0.5Z_{\sun}$, so this model should give the best age estimates. These clusters are considered to be part of the southern tail. 
\label{fig:central}}
\end{figure}

\begin{figure}
\includegraphics[width=150mm]{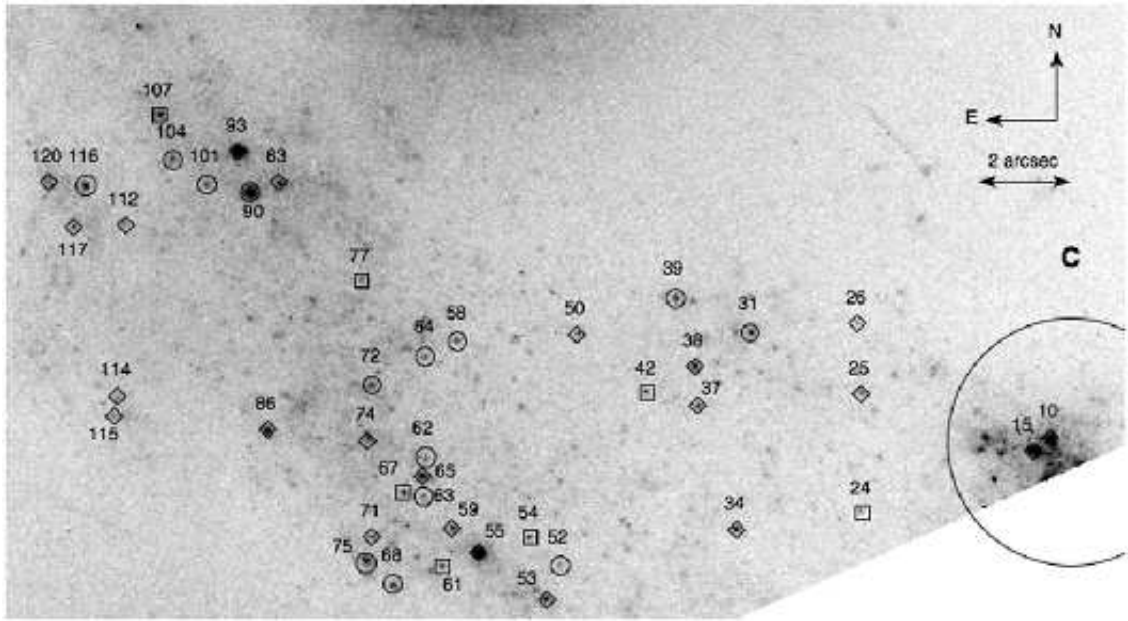}
\caption{F606W image of the southern tail of NGC 7714 with ages from the  $0.2Z_{\sun}$ models indicated for all detected clusters. Circles represent ages 1--10 Myr, diamonds 11--22 Myr, squares $\geq 23$ Myr. Clusters 10 and 15 are inside \Hii region C, which has a measured metallicity of $0.25Z_{\sun}$
\label{fig:south}}
\end{figure}

\begin{figure}
\includegraphics[width=173mm]{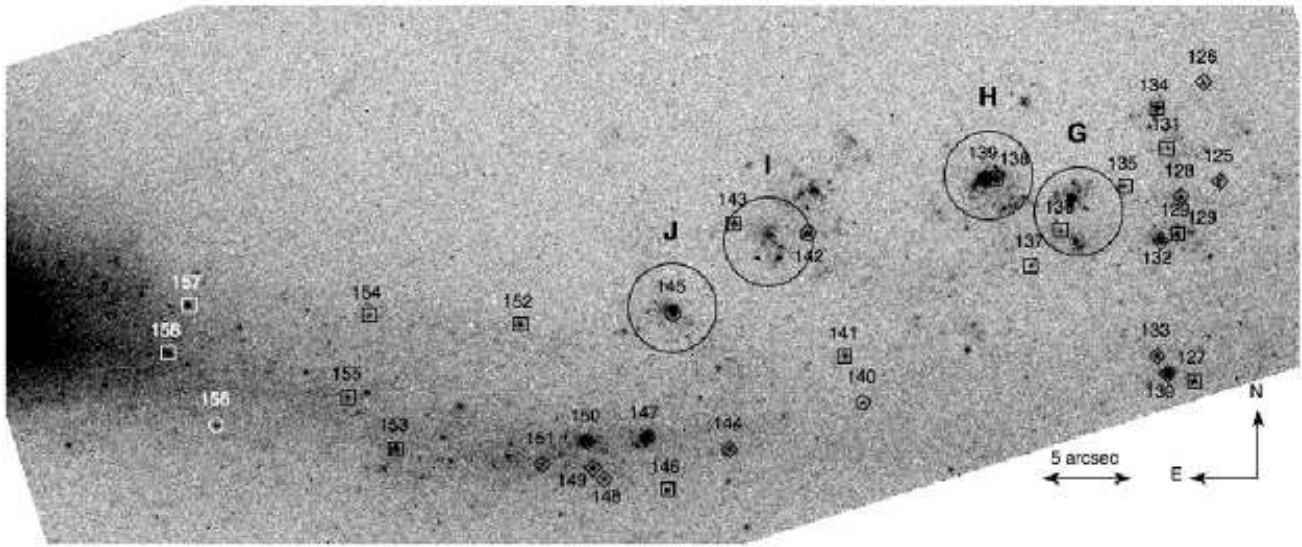}
\caption{F555W image with all detected clumps in the bridge with ages from the $0.2Z_{\sun}$ models. Circles represent ages 1--10 Myr, diamonds 11--22 Myr, squares $\geq 23$ Myr. The metallicity of the bridge has not been measured, but is likely to be low.
\label{fig:bridge}}
\end{figure}

Our best-fitting ages for the clusters are given in Table~\ref{tab:clage}, while $E(B-V)$ and masses are shown in Table~\ref{tab:clebv}. The age distributions within the features of the system are shown in Fig.~\ref{fig:ageHist}. Finding charts showing the locations of the clusters are given in Figs.~\ref{fig:regions}--\ref{fig:bridge}, with circles indicating ages $\leqslant 10$ Myr, diamonds 11--22 Myr, and squares $> 22$ Myr. The apertures used for photometry of the large \Hii regions are also shown. The clusters are generally very young, occupying the two younger age bins (1--10 and 11--22 Myr). Very few clusters older than 22 Myr are found in the tidal tails.

Ages determined by evolutionary synthesis techniques generally have large uncertainties. In our sample, the uncertainties are typically very close to the ages. The median age is 14~Myr, with a median negative uncertainty of 5~Myr and a positive uncertainty of 10~Myr.

\subsection{Luminosity function}
\begin{figure}
\includegraphics[width=84mm]{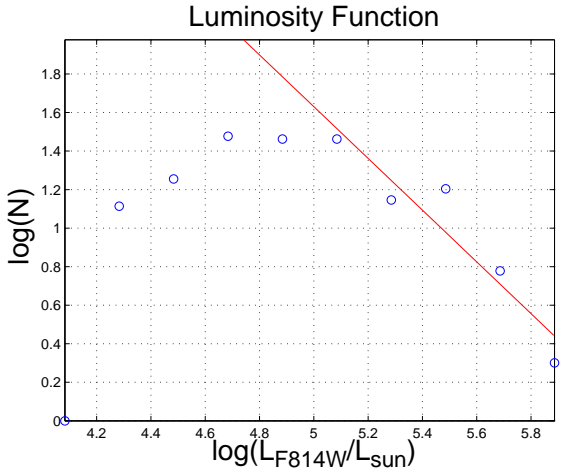}
\caption{Luminosity distribution in F814W for all detected clusters outside the nucleus. The luminosity is defined as the luminosity in the F814W band $\Delta\nu L(\nu)$, where $\Delta \nu$ is the bandwidth of $7.2\times10^{13}$ Hz. The solid red line is a power law fit to the distribution after the turnover with slope~$-1.3\pm0.3$, giving a completeness limit of approximately $(5.1\pm0.2)\times10^{5}L_{\sun}$.
\label{fig:lumfn}}
\end{figure}

The luminosities in the F814W band for the clusters $\Delta \nu L(\nu)$ were calculated using a distance of 39 Mpc (NED), using a bandwidth of $7.2\times10^{13}$ Hz. The luminosity distribution, excluding the extremely luminous nuclear population, is shown in Fig.~\ref{fig:lumfn}, and provides one estimate of our sample completeness. We assume a simple power law form for the luminosity function, defined by $n(L)dL \propto L^{-\alpha}dL$. A best-fitting line is determined after the distribution turns over, giving a slope of $-1.3$ with a completeness limit of $10^{5.1}L_{\sun}$.

The uncertainty in the slope may be estimated as follows. The distribution has a flat top, so there are two reasonable possibilities for the turnover point. Selecting as the completeness limit $10^{4.9}L_{\sun}$, we obtain a slope of slope $-1.1$.  Our best estimate for the slope is $-1.3\pm0.2$. Cluster luminosity functions typically have $\alpha = -2$ \citep[e.g.][]{whi99}, which gives a slope of $-1$ when binned logarithmically. Our luminosity function is consistent with this trend to within the uncertainties.

\section{Discussion}
\subsection{\emph{HST} cluster ages}
The metallicities determined by \citet{gon95} aid in the interpretation of our evolutionary synthesis results. In the southern tail, \Hii region A has a measured metallicity $\sim0.5Z_{\sun}$. The best age estimates should therefore come from our $0.4Z_{\sun}$ models, which show this region to be dominated by clusters with ages 10~Myr or less.

The metallicity of \Hii region C at the base of the southern tail is $\sim0.25Z_{\sun}$ \citep{gon95}. Combined with the general metallicity gradient observed in the galaxy, we are led to favor the $0.2Z_{\sun}$ model in the western part of the southern tail, which shows more clusters older than 22~Myr than does the $0.4Z_{\sun}$ model. Note, however, that part of region C was off frame (Fig.~\ref{fig:south}), so we only have age estimates for two of its constituent clusters.
	
In the northern tail, the dependence of derived age on metallicity is less pronounced. Only five clusters change age bins between the $0.2Z_{\sun}$ and $0.4Z_{\sun}$ models, all of which are near the nucleus. Most of the clusters are in \Hii regions B and D. These regions have measured abundances $\sim~0.25Z_{\sun}$, so our $0.2Z_{\sun}$ models should provide the best age estimates. The cluster ages are all less than 23~Myr in these models, with about half 10~Myr or younger. This is not unexpected, since this part of the northern tail is detected in H$\alpha$ \citep{smi97} and \emph{Spitzer} $8 \micron$ \citep{smi07}.

The metallicity of the nucleus has been measured at $\sim0.5Z_{\sun}$, so the $0.4Z_{\sun}$ model is preferred here. Seven of the 16 clusters have ages $\leq10$~Myr, with the remainder between 14--21~Myr. The gap between these age populations is probably not significant, since the rapid evolution of red giants after 10~Myr makes exact age determination exceptionally difficult and often results in gaps between 10--15~Myr \citep[e.g.][]{fall05}. The younger population fits well with the UV spectral results of \citet{gon99}, while the older population fits with one of the bursts expected by \citet{lan01}.

Of the 124 clusters detected in the tidal tails of NGC 7714, those with ages $> 22$~Myr number only five in the $0.4Z_{\sun}$ model and 12 in the $0.2Z_{\sun}$ model. If an age is assigned to each cluster based on the measured or most likely metallicity, there are eight older than 22~Myr. The small sample of 16 clusters detected in the nucleus has no clusters older than 22~Myr. There is little evidence in NGC 7714 of a population older than $\sim20$~Myr, despite an interaction that had closest approach $\sim100$~Myr ago. We also find that the cluster age groups are evenly distributed over the galaxy. The lack of older clusters detected in the nucleus may be due to the difficulty in detecting such clusters amidst the high background caused by YMCs. 

These results are similar to those obtained by \citet{bas05b} for the interacting galaxy NGC 6872. Most of the clusters in NGC 6872 were found by colour fitting to be younger than 30~Myr. These were spread throughout the galaxy, extending far out into the tidal tails. Clusters older than 30~Myr were concentrated in the central regions of the galaxy. However, the `central regions' of NGC 6872 have a spatial size $\sim30$~kpc, while the imaged portion of the NGC 7714 disc in our data is only $\sim7$~kpc. The age distribution within NGC 7714 is compatible with that found by Bastian et al. in NGC 6872, but a wider field will be required to determine if the concentration of old clusters is similar in the two galaxies.

The bridge ages are difficult to interpret because of the large colour uncertainties, incomplete coverage in F380W, and lack of metallicity data. The metallicity in NGC 7714 generally decreases with distance from the nucleus. However, some of the bridge material may originate in NGC 7715, which has unknown metallicity. Nevertheless, a low metallicity is expected due to the abundances measured elsewhere in the system and because tidal features in interacting galaxies generally show low abundances.

Despite the large uncertainties, we note that the only evidence for a significant population with age $\ga100$ Myr is in the bridge. Out of 34 clusters, the numbers with ages $> 22$~Myr are 12 and 17 in the $0.4Z_{\sun}$ and $0.2Z_{\sun}$ models, respectively. However, only one of these clusters has F380W data available in the $0.4Z_{\sun}$ case, and nine in the $0.2Z_{\sun}$ case. The existence of an older population in the bridge is by no means certain, but if present could indicate that cluster dissolution mechanisms have acted more weakly here than in the tidal tails, which in this system are closer to the nucleus. Older clusters should survive more easily in the bridge because tidal forces are generally smaller there than in the tails. However, the gas expulsion mechanisms thought to be responsible for infant mortality operate independently from the tidal forces, so the rate at which very young clusters dissolve is not expected to change.
	
To address the issue of sample completeness, we begin with the luminosity function (Fig.~\ref{fig:lumfn}). Cutting out clusters with luminosities below the completeness limit, we have 49 left in the tidal tails and only four in the bridge. The bridge sample is far too incomplete to permit any reliable analysis.
	
Only two clusters in the tidal tails that survive the luminosity cut have ages $> 22$ Myr, and those only in the $0.2Z_{\sun}$ model.  It is unsurprising that older clusters would be preferentially cut from a luminosity-limited sample, since clusters get both redder and fainter as their stellar populations age.

\subsection{\emph{HST} cluster mass function} \label{sec:mass}
It is common practice in studies of cluster ages to use a mass cutoff rather than a luminosity cutoff. Due to evolutionary fading, clusters of a given age will remain above the luminosity limit only if they have a sufficiently high mass. Therefore, imposing a luminosity limit will generally under-sample old clusters.
	
Mass estimates are based on the assumption that clusters of a given age all have the same mass to light ratio, with luminosity values calculated for clusters with mass $10^6M_{\sun}$. The luminosity of a cluster relative to a $10^6M_{\sun}$ cluster of the same age then gives an estimate of the mass. The uncertainties in the masses are quite large, up to a factor $\sim2$. Mass estimates based on the best-fitting age and extinction are shown in Table~\ref{tab:clebv}.

\begin{figure}
\includegraphics[width=84mm]{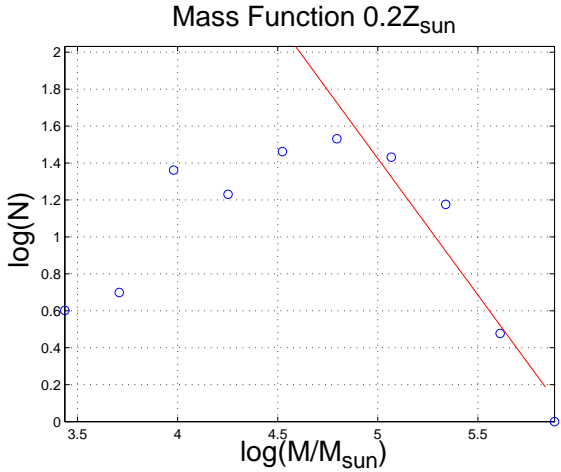}
\caption{Cluster mass distributions for the $0.2Z_{\sun}$ model. The solid red line is a power law fit to the part of the distribution after the turnover. The fit has slope $-1.5\pm0.4$ with a limiting mass $\sim10^{4.8}M_{\sun}$.
\label{fig:mass}}
\end{figure}

The lowest mass for which the sample is complete was determined assuming a simple power law form for the cluster mass distribution. For $0.4Z_{\sun}$, we obtain a reasonably good fit with a slope $\approx-1.4$ and a limiting mass of $10^{5.0}M_{\sun}$. The quality of the fit is not high due to a lack of clusters in the mass bin centered at $10^{6.0}M_{\sun}$, which may reflect the high uncertainties in masses determined in this way. The mass estimates for the $0.2Z_{\sun}$ model produced a slope of $-1.5$, but with a lower mass limit at about  $10^{4.8}M_{\sun}$. The mass distribution for the $0.2Z_{\sun}$ model is shown in Fig.~\ref{fig:mass}. 

The uncertainties for the mass function fits may be estimated using the same technique as was used for the luminosity function. For the $0.4Z_{\sun}$ model, the next best choice of limiting mass would be $10^{4.7}M_{\sun}$, which gives a slope of of $-1.0$. In the $0.2Z_{\sun}$ model, both $10^{4.5}M_{\sun}$ and $10^{5.0}M_{\sun}$ are reasonable choices, producing slopes of $-1.8$ and $-1.1$, respectively. Using these alternate choices to estimate the uncertainties, the $0.4Z_{\sun}$ model has a slope $-1.4\pm0.4$ , with smaller values being more likely. For the $0.2Z_{\sun}$ model, the slope is $-1.5\pm0.4$. At the limits of the uncertainties, our mass functions agree with the typical power law index near $\alpha=-2$, found for example in the Antennae \citep{zha99} and Magellanic Clouds \citep{hun03}. Due to the small size of our sample, we are unable to test the claim of \citet{gie09} that the cluster initial mass function is truncated at the high mass end.

\begin{figure}
\includegraphics[width=84mm]{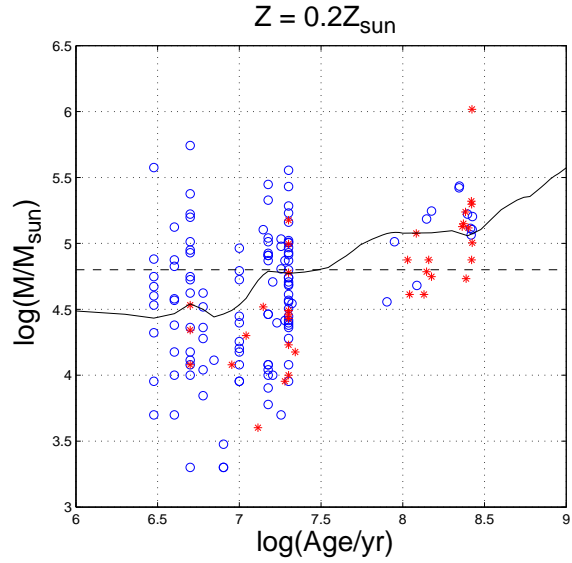}
\caption{Mass vs. Age for the $0.2Z_{\sun}$ model. Blue circles represent the NGC 7714 disc, while red asterisks are bridge clusters. The dashed black line represents the age-averaged mass completeness limit of $10^{4.8}M_{\sun}$ from Fig.~\ref{fig:mass}. The solid black curve shows the mass of a cluster with an F840W luminosity equal to our completeness limit of $10^{5.1}L_{\sun}$, as a function of age.
\label{fig:MvAge}}
\end{figure}

The completeness limit in mass, unlike that in luminosity, is a function of age due to evolutionary fading. Thus the turn-over in Fig.~\ref{fig:mass} gives an age-averaged completeness limit for mass. For younger systems, the mass completeness limit is lower than for older systems. The masses are plotted against the ages for the $0.2Z_{\sun}$ model in Fig.~\ref{fig:MvAge}, with the NGC 7714 disc clusters as blue circles and bridge clusters as red asterisks. The solid black curve shows the mass of a cluster with an F840W luminosity equal to our completeness limit of $10^{5.1}L_{\sun}$, as a function of age. At lower ages, we are reaching less massive clusters, and are more deficient at older ages. For clusters less than 10 Myr old, we are complete to about $10^{4.3}M_{\sun}$, while for $10^{9}$ yr we are complete only to $10^{5.5}M_{\sun}$. The dashed black line indicates the average completeness limit of $10^{4.8}M_{\sun}$. The significance of evolutionary fading is apparent, as all of the clusters with ages $\sim10^8$ yr have masses $\geq10^{4.5}M_{\sun}$. 

Mass cuts alter the age distributions considerably. In the $0.4Z_{\sun}$ model, seven of the 27 clusters remaining after eliminating those with mass $< 10^{5.0}M_{\sun}$ are older than 22 Myr, with the youngest of these 111 Myr old. These account for six of the seven clusters remaining in the bridge (only one of which has F380W coverage) and one of six in the northern tail, but none of the 14 clusters in the southern tail.

Of the 63 clusters remaining after eliminating those with mass $< 10^{4.8}M_{\sun}$ in the $0.2Z_{\sun}$ model, 31 are in the southern tail, 18 in the northern tail, and 14 in the bridge. There are 22 clusters with ages $>22$ Myr, 12 of them in the bridge, of which five lacked coverage in F380W. In the southern tail, seven of 31 clusters are older than 22 Myr (five of them detected at $3\sigma$ level), while the northern tail has two of 18 clusters older than 22 Myr. The youngest of these clusters has an age of 89 Myr.

The number of clusters falling into each age bin in the $0.2Z_{\sun}$ model is summarized in Table~\ref{tab:bin02} for all measured clusters, clusters above the luminosity limit, and clusters above the average mass completeness limit. For comparison, we have included columns showing the expected numbers of clusters assuming a cluster formation rate of 1 cluster Myr$^{-1}$, and constant cluster formation for the last $10^{7.5}$ (CCF$_{30}$) and $10^8$(CCF$_{100}$) yr, only including clusters above the mass limits. These are limiting cases, based on Fig.~\ref{fig:mass}. The observed number of clusters in the mass limit case fall between these two limits. Thus with the available data we cannot rule out constant cluster formation with no infant mortality. More sensitive data are needed to reduce the completeness limits and search for evidence of cluster destruction. With the current data set, the number of clusters is relatively small and the mass cut-off relatively high, thus the results on infant mortality are uncertain.

\subsection{\Hii region age estimates} \label{sec:hiiage} 
Ages for the \Hii regions were estimated in two ways. The colours obtained from large-aperture photometry in the \emph{GALEX} and \emph{HST} bands were fit to {\sc sb99} models, using the methods described in \S~\ref{sec:aged}. In regions E and G--J, the FUV and NUV measurements provided only upper limits, so these bands were not fit to the models. The range of cluster ages within the \Hii regions suggests that the star formation histories are more complex than the instantaneous starbursts used to model the individual clusters. We therefore used both instantaneous and continuous star formation models to estimate the ages of these regions. The true star formation history is likely to fall between these two extremes.

Ages were also estimated using the H$\alpha$ equivalent widths, which were fit to instantaneous burst {\sc sb99} models. These data are shown in Table~5, along with the median age of the \emph{HST} clusters detected in each region. Note that the \emph{HST} sample has no clusters in region E (four  were detected but were covered in only two filters), only two clusters in region C, since it is at the edge of the WFPC2 field of view, and in the bridge (regions G--J) only one or two clusters were found in each region, so this indicator is only useful in regions A, B, D, and F. When possible, we also compare with published ages determined via photoionisation modeling.

The ages determined using EW(H$\alpha$) are generally consistent with the median age of clusters found in the region, with both showing cluster population ages $\la15$ Myr. These estimates are somewhat higher than those in the literature (see Table~5), but are within a factor of $\sim3$ in most of the large aperture photometry cases.

The ages determined by fitting {\sc sb99} models to the large aperture photometry of the \Hii regions are much higher than those determined by other methods in regions B, D, and F. The area within the apertures occupied by clusters is relatively small, so it is possible that faint sources outside the clusters are making significant contributions to the total luminosity of the \Hii regions. If these sources are older and redder, the age of the region as a whole could be quite different from that of its bright clusters. These sources could be older clusters below our detection threshold or the remnants of clusters that have already dissolved and entered the field population.

To quantify the possible influence of faint sources on the \emph{HST} apertures, the fraction of luminosity in the F814W band for each of the \Hii regions $f_{c}$ originating in the measured clusters was determined. The $f_{c}$ values are similar throughout the NGC 7714 disc, with $\sim80$ per cent of the light coming from outside the clusters. Unresolved objects could significantly impact the colours, which may explain the higher age estimates for regions B, D, and F.

All of the bridge \Hii regions had their lowest age estimates provided by colour fitting, while $f_{c}$ varies. Regions G and I had the lowest $f_{c}$ values in the system, while H and J have the highest. Visual inspection of regions G and I indicates several likely clusters that were not selected by \emph{daofind}. Considering the relatively high noise and low resolution in \emph{HST} bridge data, it is likely that these results are products of the data quality. Deeper imaging will be required to learn if the bridge hosts a cluster population that differs from that of the NGC 7714 disc.

Extinction estimates are shown in Table~\ref{tab:hiiebv}, along with some values from the literature. The extinction estimates of \citet{gon95} come from the Balmer decrement, while the extinction in the nucleus found by \citet{gon99} was determined using evolutionary synthesis results for UV continuum flux distributions, and was smaller than that found using the Balmer decrement by a factor $\sim3$. Our extinction estimate for the nucleus is in good agreement with that of \citet{gon99}, while our other estimates are higher than the literature values by a factor $\sim2$. This is not surprising, since the Balmer decrement tends to give higher extinction estimates than the continuum by a factor $\sim2$ \citep{fan88, sto94}, possibly because dust in star forming regions with strong nebular emission is destroyed or removed by ionizing radiation, stellar winds, or supernovae \citep{fan88}. Another possibility is that the hot stars responsible for the Balmer lines are still associated with the dusty molecular clouds in which they formed, while the cooler stars that contribute to the continuum have drifted away from these regions and suffer less extinction \citep{cal94}.

\subsection{\emph{Spitzer} colours} \label{sec:spitz}
\begin{figure}
\includegraphics[width=84mm]{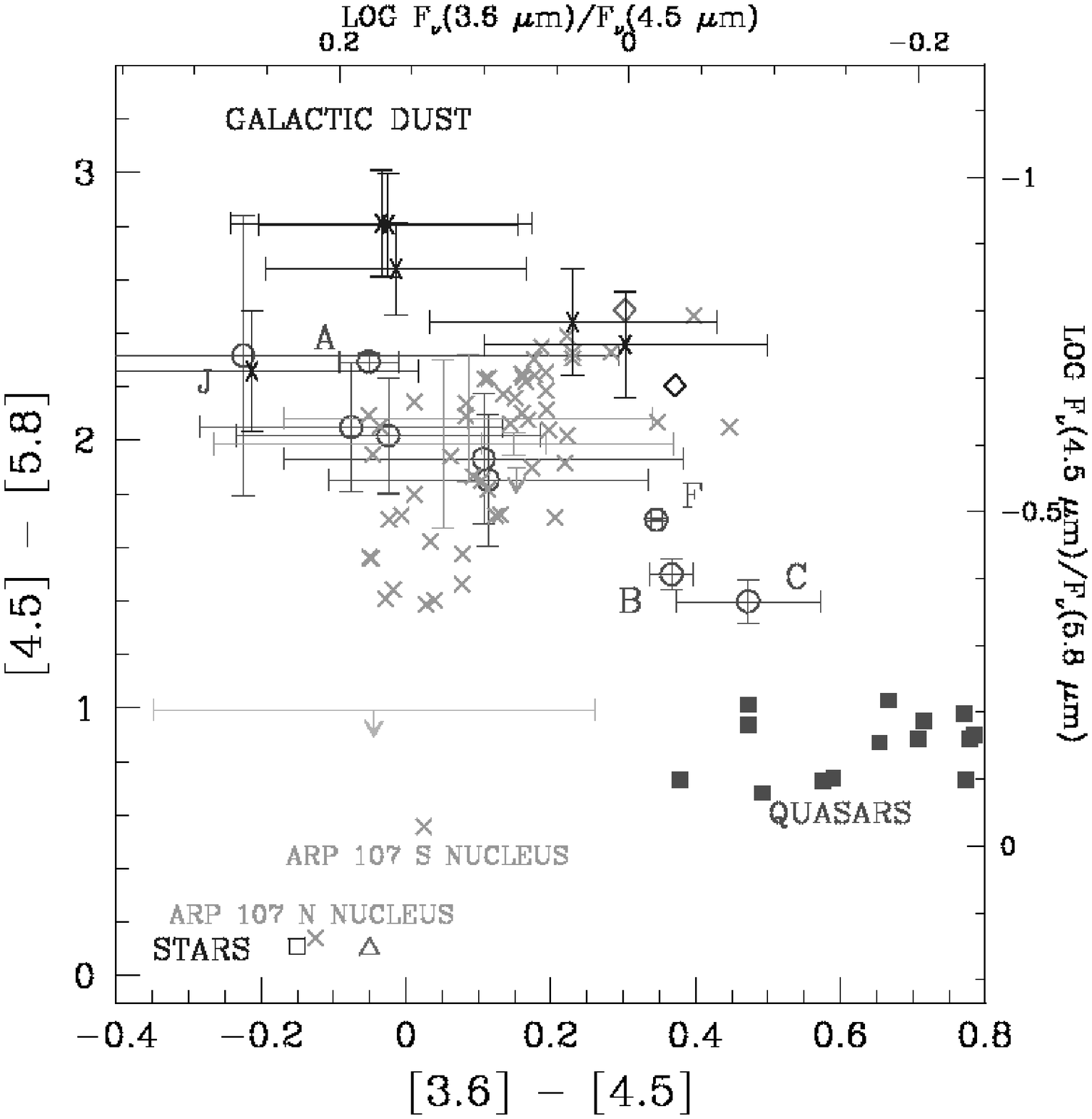}
\caption{\emph{Spitzer} [4.5]$-$[5.8] vs. [3.6]$-$[4.5] colour-colour plot. The Arp 284 \Hii regions (open red circles) are shown along with emission `clumps' from Arp 24 \citep{cao07}, 82 \citep{han07}, 107 \citep{smi05b}, 285 \citep{smi08}, and NGC 2207/IC 2163 (\citealt{elm06}; green crosses). Shown separately are Arp 285 NGC 2856 tail clump 3 (magenta open diamond) and disc clump 1 (cyan limit), as well as NGC 2207/IC 2163 `clump i' shown separately (open blue diamond). Also shown are the mean colours of field stars of \citet{whi04} (magenta open triangle), M0III stars (M. Cohen 2005, private communication; open blue square), quasars (\citealt{hat05}; red squares), and diffuse dust in the Milky Way (\citealt{fla06}; blue crosses).
\label{fig:spi1223}}
\end{figure}

\begin{figure}
\includegraphics[width=84mm]{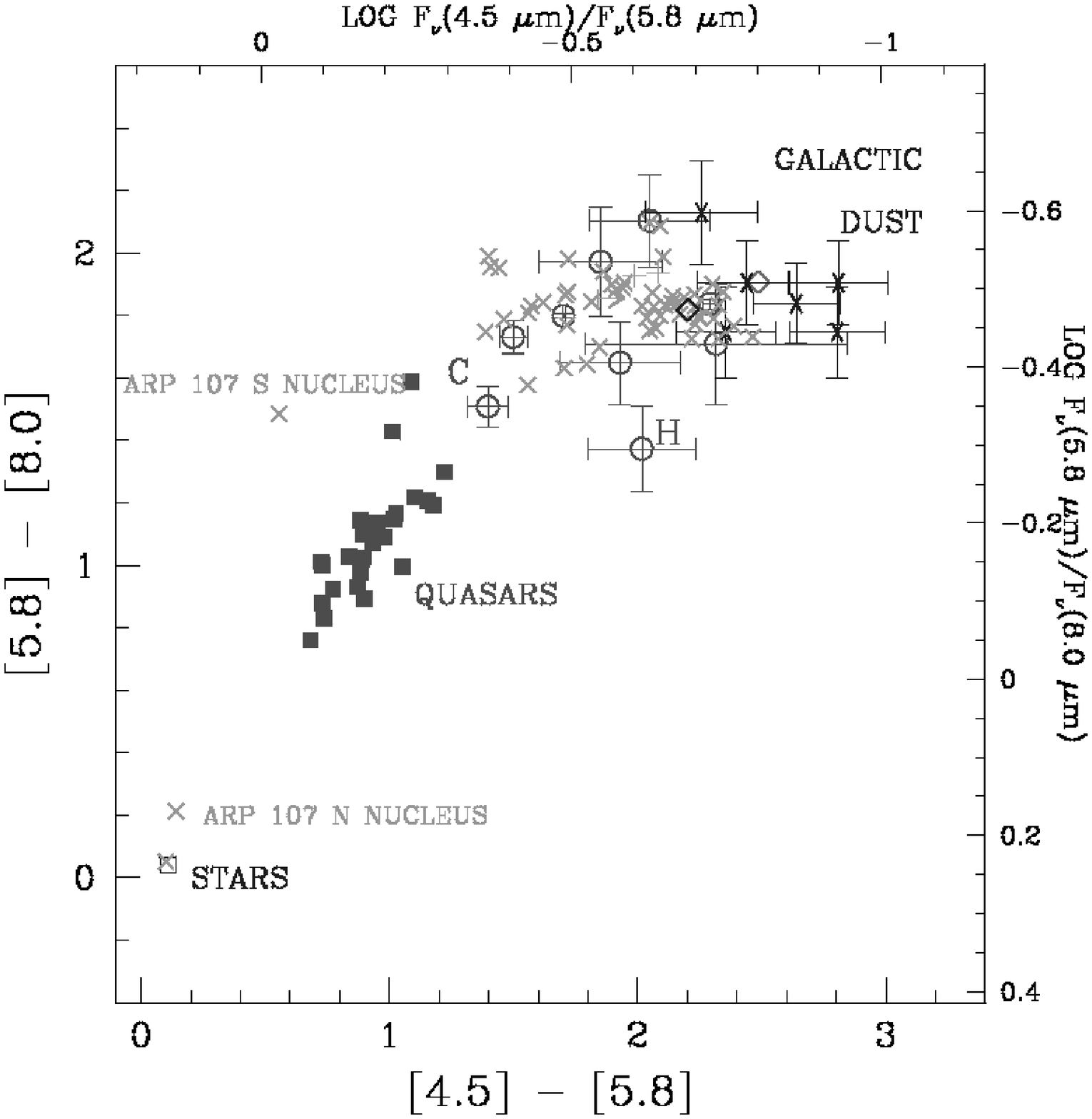}
\caption{Same as Fig.~\ref{fig:spi1223}, but for [5.8]$-$[8.0] vs. [4.5]$-$[5.8].
\label{fig:spi2334}}
\end{figure}

Figs.~\ref{fig:spi1223} and \ref{fig:spi2334} show as red circles the \emph{Spitzer} [4.5]$-$[5.8] vs. [3.6]$-$[4.5] and [5.8]$-$[8.0] vs. [4.5]$-$[5.8] colours of the \Hii regions. These were calculated from the flux densities as described in the IRAC Data Handbook. For comparison, we also show emission `clumps' from several other interacting systems, including Arp 24 \citep{cao07}, 82 \citep{han07}, 107 \citep{smi05b}, 285 \citep{smi08} and NGC 2207/IC 2163 \citep{elm06}, as green crosses. In the Arp 285 system, NGC 2856 tail clump~3 and disc clump~1 were unusual, and are marked in Fig.~\ref{fig:spi1223} as a magenta open diamond and cyan limit, respectively. The unusually luminous clump `i' in the eastern tail of NGC 2207 is also shown separately, as an open blue diamond. We also display the mean colours of field stars of \citet{whi04} (magenta open triangle), M0III stars (M. Cohen 2005, private communication; open blue square), quasars (\citealt{hat05}; red squares), and diffuse galactic dust (\citealt{fla06}; blue crosses).

Most of the \Hii region \emph{Spitzer} colours fit well with the clumps measured in other systems. This is important, since the selection criteria are somewhat different. The Arp 284 regions were selected based on H$\alpha$ emission, whereas clumps in the other systems were selected based on \emph{Spitzer} emission. While the H$\alpha$ emission usually had a counterpart in \emph{Spitzer} $8.0\micron$, in some cases the aperture placement would have differed slightly if the $8.0\micron$ emission had been used for selection.

[4.5]$-$[5.8] is often used as a gauge of the star formation rate, with redder colours indicating higher rates of mass-normalized star formation \citep[e.g.][]{smi05b}. The \Hii regions lie between interstellar matter and stars, indicating that, as in other interacting systems, they probably have contributions from both. However, none of the Arp 284 regions is quite as red as the Arp 285 tail clump. \citet{smi08} suggest that the exceptionally red colour of this clump is due to strong emission from interstellar matter owing to the very young age of the clump, which was determined to be 4 Myr using optical colours. Region J has the same estimated age and is only slightly bluer in colour, suggesting that it may also have numerous hot young stars.

The [5.8]$-$[8.0] colours generally fall close to those of interstellar matter, which is not surprising since the 5.8 and 8.0$\micron$ bands are expected to be dominated by interstellar dust.

A few of the regions are unusual. Regions B and C are unusually red in [3.6]$-$[4.5], but slightly blue in the other two colours. Region F is redder in [3.6]$-$[4.5] than clumps in other systems with similar [4.5]$-$[5.8] colour. Region H is slightly blue in [5.8]$-$[8.0].

\begin{figure}
\includegraphics[width=84mm]{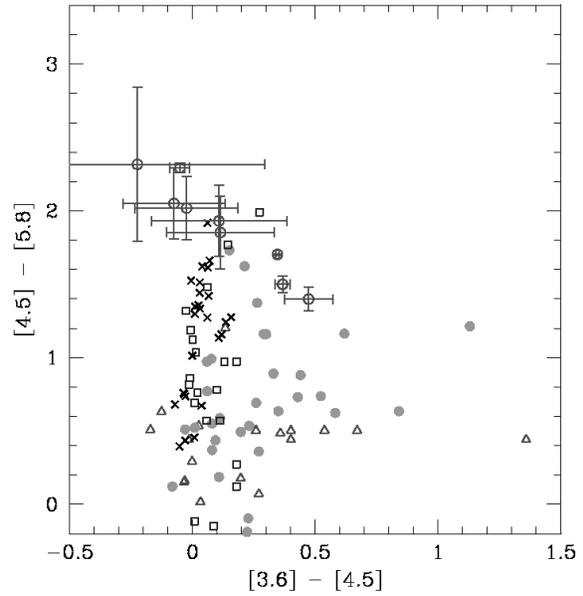}
\caption{\emph{Spitzer} [4.5]$-$[5.8] vs. [3.6]$-$[4.5] colour-colour plot showing Arp 284 \Hii regions (open red circles). The dwarf galaxies of \citet{smi09} are shown by metallicity, with $\mathrm{12+\log(O/H)<7.9}$ (red open triangles), $\mathrm{7.8\leqslant 12+ \log(O/H) \leqslant 7.9}$ (green filled circles), and $\mathrm{12+\log(O/H)>8.2}$ (blue open squares), along with a sample of `normal' spirals (black crosses). 
\label{fig:spid1223}}
\end{figure}

\begin{figure}
\includegraphics[width=84mm]{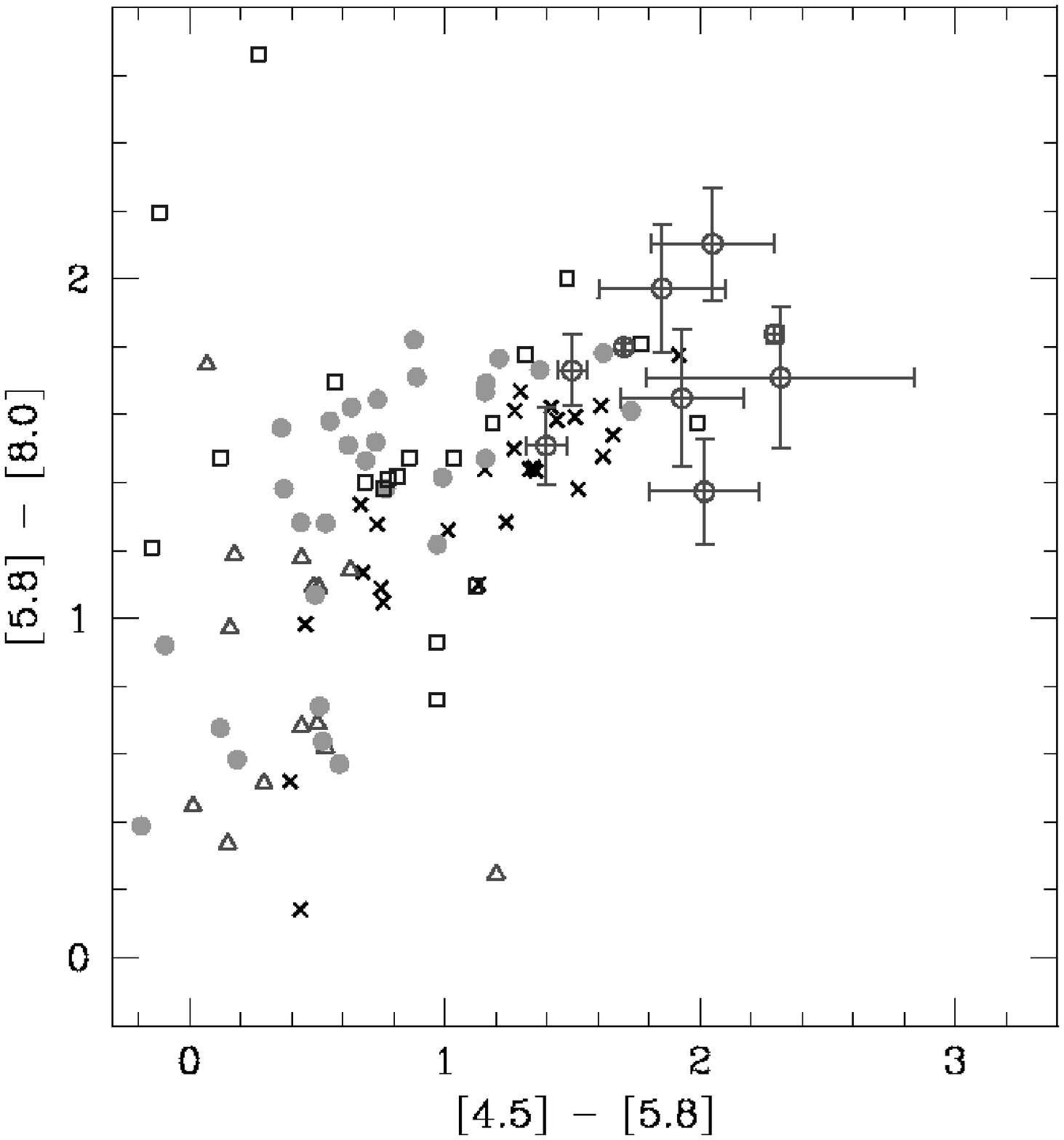}
\caption{Same as Fig.~\ref{fig:spid1223}, but for [5.8]$-$[8.0] vs. [4.5]$-$[5.8].
\label{fig:spid2334}}
\end{figure}

The colours of regions B and C suggest excess emission at $4.5\micron$, perhaps with a slight deficiency at 5.8 and $8.0\micron$, qualitatively similar to low-metallicity dwarf galaxies. In Figs.~\ref{fig:spid1223} and \ref{fig:spid2334}, we compare the \emph{Spitzer} colours of Arp 284 \Hii regions to those of the dwarf and spiral galaxies of \citet{smi09}. Regions B and C are not as red in  [3.6]$-$[4.5] as some the dwarfs, but are generally redder than the spirals. This may indicate that they are deficient in PAHs, though not extremely so, consistent with their moderately low metallicities.

It is interesting that regions B and C stand out in these plots, but the other regions do not. Our colour-based age estimate for region B is very similar to that of D, and our estimates based on EW(H$\alpha$) vary little between regions. \citet{gar97}, however, found regions B and C to be extremely young. This would imply more nebular continuum and Br$\alpha$ emission, which are two of the factors that appear to contribute to reddening [3.6]$-$[4.5] in the dwarfs \citep{smi09}.

\begin{figure}
\includegraphics[width=84mm]{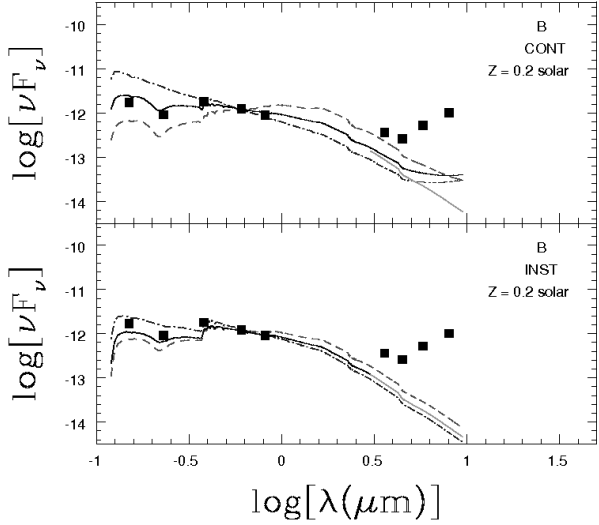}
\caption{The FUV-mid-IR SED of region B. The black curves are the best-fitting $0.2Z_{\sun}$ evolutionary synthesis models including both stellar and nebular emission. The model in the top panel is for continuous star formation with an age of $128^{+1272}_{-42}$ Myr, while the bottom panel shows an instantaneous burst  with an age of $88^{+16}_{-18}$ Myr. The solid green line plotted from 3--9.5$\micron$ shows the models without nebular emission. The short-dashed red and dot-dashed blue curves correspond to the youngest and oldest models and the associated best-fitting extinctions. The models have been scaled to match the observations in F606W.
\label{fig:sedB}}
\end{figure}

The SED for region B is plotted in Fig.~\ref{fig:sedB}. We also show as black curves the best-fitting $0.2Z_{\sun}$ evolutionary synthesis models from {\sc sb99} for both continuous (top panel) and instantaneous burst (bottom panel) star formation models. The curves use the best-fitting ages for each model, which are $128^{+1272}_{-42}$ Myr for continuous star formation and $88^{+16}_{-18}$ Myr for the instantaneous star formation model. The models include both stellar and nebular emission. The solid green lines show the same models without nebular emission. The models corresponding to 1$\sigma$ uncertainty (68 per cent confidence) are also shown. All of the curves have been scaled to match the observations in F606W. The stellar and stellar+nebular emission models are nearly indistinguishable, indicating that nebular emission is relatively unimportant for these inferred ages. The large mid-IR excess could result from hot dust continuum emission, PAH lines, or  an older stellar population, either in clusters that have faded below our detection limits or in the field. The excess is diminished in the continuous star formation model, probably because it includes older stars.

\begin{figure}
\includegraphics[width=84mm]{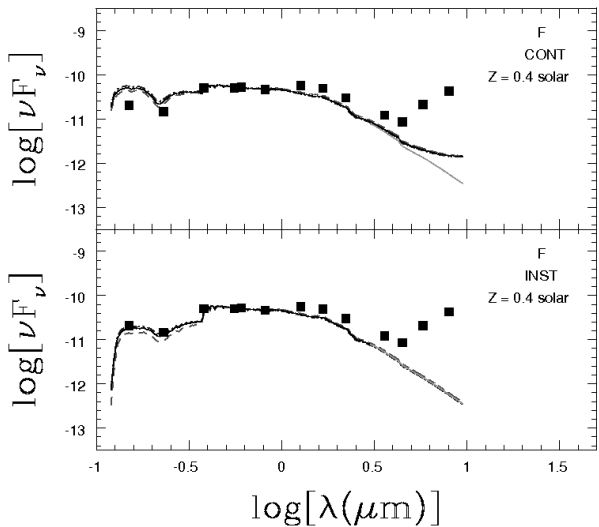}
\caption{Same as Fig.~\ref{fig:sedB}, but shows region F with $0.4Z_{\sun}$ models, with ages of $421^{+12}_{-11}$ Myr for the continuous star formation model and $222^{+6}_{-6}$ Myr for the instantaneous burst model.
\label{fig:sedF}}
\end{figure}

For comparison, the SED of the nucleus is shown in Fig.~\ref{fig:sedF}. The top panel shows curves for a continuous star formation model with an age of $421^{+12}_{-11}$ Myr, while the bottom panel shows an instantaneous starburst with an age of $222^{+6}_{-6}$ Myr. In this plot we have also included the near-IR \emph{J}, \emph{H}, and \emph{K} band fluxes from \citet{lan01}. The nucleus is known to have a complex star formation history, likely including stars older than $\sim100$~Myr. The mid-IR excess is greater in the instantaneous burst model, likely due older stars not included in this model.

\subsection{Comparison with \emph{Chandra}}
\subsubsection{Ultra-luminous X-ray sources (ULXs)}
X-ray observations by \citet{smi05a} detected three non-nuclear point sources with $L_{X} > 10^{39}$ erg s$^{-1}$, the standard definition of an ultraluminous X-ray (ULX) source. Three other point sources lie just below this limit. The nature of ULXs is not presently understood. In one possible scenario, they could be intermediate mass ($100-1000M_{\sun}$) black holes (IMBH; \citealt{col99}). Alternatively, they may be stellar mass black holes with beamed \citep{kin01} or super-Eddington \citep{beg02} X-ray emission. IMBHs may preferentially form in massive star clusters \citep[e.g.][]{por04b}. It takes at least 4 Myr for a stellar mass black hole/X-ray binary to form, and they are expected to be common in populations with ages up to $\approx 100$ Myr \citep{rap04}. In contrast, an IMBH may form in a YMC in $\le 3$ Myr \citep{por04b}. Evolutionary models of the predicted X-ray emission from IMBH+stellar companion mass-transfer binaries indicate that such ULXs should generally have ages less than about 15--30 Myr \citep{por04a}.

As discussed in \citet{smi05a}, the ULX candidates in NGC 7714/5 exist in a range of environments. A luminous X-ray point source is located $1.5\arcsec$ away from the nucleus, near cluster n01, which has an estimated age of 14 Myr. If this association is correct, the young age is more consistent with the IMBH scenario. Another possibly young ULX candidate is found within \Hii region C, which has an age of 11 Myr based on EW(H$\alpha$) and a spectroscopically determined age of 4.5 Myr \citep{gar97}. However, this second ULX cannot be associated with a specific star cluster, since there are several very close together in the area. Finally, one ULX is located in the outer western tail of NGC 7714, in an area that is undetected in H$\alpha$ but is bright in \emph{GALEX} images. Three other ULX candidates are neither located in \Hii regions nor are they close to optically-selected star clusters. Given the variety of environments, from this small sample we cannot make any strong conclusions about the nature of ULXs as a class.

\subsubsection{Extended X-ray emission associated with star forming regions}
In addition to point sources, the \emph{Chandra} map revealed extended X-ray emission associated with some of the \Hii regions \citep{smi05a}. Fig.~\ref{fig:chandra} shows \emph{Chandra} 0.3--8 keV X-ray contours over a colour scale H$\alpha$ image of NGC 7714. Extended X-ray emission is found associated with all of the disc \Hii regions except D. The bridge regions G--J also do not have observed diffuse X-ray emission.

The possible contributors to the extended X-ray emission include multiple unresolved high mass X-ray binaries (HMXBs), or hot gas from supernovae and stellar winds impacting the ambient interstellar medium. Based on Local Group studies, the HMXB component is expected to be too small to account for this emission in the NGC 7714/5 regions \citep{smi05a}, unless there is an excess above Local Group galaxies. For the hot gas component, using {\sc sb99} models and a constant X-ray production efficiency $L_X$/$L_{mech}$, where $L_{mech}$ is the mechanical luminosity from the supernovae and winds, the ratio of the number of Lyman continuum photons $N_{LyC}$ to X-ray luminosity $N_{LyC}$/$L_X$ is expected to drop off with age for a star forming region \citep{smi05a}. However, the one \Hii region complex in the NGC 7714 disc that is not detected in extended X-ray emission, region D, has an EW(H$\alpha$)-based age that is somewhat older than some of the other regions (Table~\ref{tab:hiiage}), in spite of having the highest $N_{LyC}$/$L_X$. In contrast, region E is bright in the X-rays, with a low $N_{LyC}$/$L_X$, but has a younger age from its EW(H$\alpha$). This suggests that the X-ray production efficiency may vary with time and/or from region to region, and/or the HMXB contribution may be larger than expected in some regions. This result is uncertain, however, due to the relatively large uncertainties on our age estimates and the small number of \Hii regions in our sample.

\section{Conclusions}
We have used broadband colours to obtain age estimates for 174 proto-globular cluster candidates in the Arp 284 system. The populations detected in the tidal features of NGC 7714 are generally quite young ($\la20$ Myr old). However, when the sample is limited by a mass cut, the dominance of young clusters becomes unclear in the NGC 7714 disc and vanishes in the bridge. Due to the small number of clusters above the mass cut, this latter result is highly uncertain. Deeper imaging is required to better define the cluster mass function and push down the completeness limit, particularly in the bridge, which has higher photometric uncertainties and less F380W coverage than the rest of the system.

No substantial differences are found between the populations within NGC 7714. The clusters in the nucleus do not generally differ in age from those in the tidal structures. It therefore appears that the early-stage dissolution processes, if present, operate via internal processes, rather than local environmental effects, but we cannot rule out the possibility of weaker dissolution in the bridge.

Using larger apertures to study \Hii region complexes, we find in several cases that the region as a whole appears to be older than the clusters detected inside it. This suggests an older, redder population that is unresolved, possibly the remnants of star clusters that have already dissolved or which were not detected as discrete sources because of evolutionary fading. 

The large aperture results suggest that star formation has been ongoing in \Hii regions B and D, located in the northern tail, for a dynamical time-scale, like that in the nucleus. In both cases, the star formation history, whether continuous or in discrete bursts, is unknown. We also do not know if regions B and D are gravitationally bound, or simply regions where successive waves or infall events have triggered star formation. Region C is similar, but we have insufficient data to comment on its nature. If in the future these regions are found to be bound, they could be interpreted as nearby analogs to the high redshift disc clumps studied by \citet{elm09}.

The Arp 284 interaction has been in progress for over 100 Myr, yet we find little evidence for a significant cluster population older than 20 Myr. Thus it is possible that clusters are dissolving on a relatively short time-scale. However, due to completeness issues, with the available data we cannot definitively confirm the possibility of infant mortality in this system. Further, more sensitive observations are needed to search for a deficiency of older clusters.

\section*{Acknowledgments}
We thank the \emph{HST}, \emph{Spitzer}, \emph{GALEX}, and \emph{Chandra} teams for making this research possible. We also thank Nate Bastian and the anonymous referee for helpful comments and suggestions. We acknowledge support from NASA Spitzer grant 1347980 and NASA Chandra grant AR90010B. BJS acknowledges support from NASA LTSA grant NAG5-13079. This research has made use of the NASA/IPAC Extragalactic Database (NED) which is operated by the Jet Propulsion Laboratory, California Institute of Technology, under contract with the National Aeronautics and Space Administration.

\bsp    


\clearpage
\newpage

\clearpage
\newpage

\clearpage
\newpage

\clearpage
\newpage

\clearpage
\newpage

\clearpage
\newpage

\clearpage
\newpage
\begin{table}
\caption{\emph{HST} WFPC2 imaging of NGC 7714/5}
\label{tab:hst}
\begin{tabular}{ccrrrrrrrrrrr}
\hline
\hline
\multicolumn{1}{c}{Dataset} &
\multicolumn{1}{c}{Exposure} &
\multicolumn{1}{c}{Filter} &
\multicolumn{1}{c}{\Hii Regions}&
\multicolumn{1}{c}{\Hii Regions}\\
\multicolumn{1}{c}{} &
\multicolumn{1}{c}{(sec)} &
\multicolumn{1}{c}{} &
\multicolumn{1}{c}{in PC chip}&
\multicolumn{1}{c}{in WF chip}&
\multicolumn{1}{c}{Comment}\\
\hline
U3GQ0201R&500&F380W&nucleus,A,B,part C,D,E&bridge & bridge split WF3/4\\
U3GQ0202R&500&F380W&nucleus,A,B,part C,D,E&bridge & bridge split WF3/4\\
U3GQ0203R&400&F380W&nucleus,A,B,part C,D,E&bridge&CR on PC chip\\
 & & & & & bridge split WF3/4 \\
U3GQ0204M&400&F380W&nucleus,A,B,part C,D,E&bridge & bridge split WF3/4\\
U6A01601R&350&F555W&nucleus,A,E&bridge\\ 
U6A01602R&350&F555W&nucleus,A,E&bridge\\ 
U2E68801T&500&F606W&nucleus,A,B,part C,D,E&bridge\\
U6A01701R&350&F814W&nucleus,A &B,C,part D\\
U6A01702R&350&F814W&nucleus,A &B,C,part D\\
U6A01801R&350&F814W&nucleus,A, E &bridge\\
U6A01802R&350&F814W&nucleus,A, E &bridge&CR on PC chip\\
\hline

\end{tabular} 
\end{table}
\begin{table}
\caption{Cluster magnitudes and age estimates. The full table is available in the electronic edition of the journal.} 
\label{tab:clage}
\begin{tabular}{cccclrrr} 
\hline \hline 
ID & F380W& F555W & F606W & F814W & Age ($Z_{\sun}$) & Age ($0.4Z_{\sun}$) & Age ($0.2Z_{\sun}$)\\ 
  & mag & mag & mag & mag & Myr & Myr & Myr\\ 
\hline
1 & 22.46$\pm$0.06 & -- & 22.64$\pm$0.03 & 22.18$\pm$0.05 & 15$^{+1}_{-1}$ & 15$^{+3}_{-6}$ & 15$^{+3}_{-1}$ \\
2 & 23.30$\pm$0.14 & -- & 22.84$\pm$0.06 & 23.10$\pm$0.15 & 4$^{+2}_{-2}$ & 5$^{+2}_{-2}$ & 5$^{+4}_{-1}$ \\
3 & 22.33$\pm$0.07 & -- & 21.23$\pm$0.03 & 21.67$\pm$0.07 & 1$^{+1}_{-1}$ & 2$^{+2}_{-2}$ & 3$^{+1}_{-2}$ \\
4 & 22.46$\pm$0.08 & -- & 22.02$\pm$0.04 & 22.09$\pm$0.08 & 6$^{+1}_{-2}$ & 6$^{+1}_{-1}$ & 7$^{+8}_{-2}$ \\
5 & 21.25$\pm$0.04 & -- & 20.86$\pm$0.02 & 21.06$\pm$0.04 & 5$^{+1}_{-1}$ & 5$^{+1}_{-1}$ & 6$^{+1}_{-1}$ \\
6 & 22.18$\pm$0.06 & -- & 22.04$\pm$0.03 & 22.28$\pm$0.09 & 5$^{+1}_{-2}$ & 5$^{+1}_{-1}$ & 6$^{+1}_{-1}$ \\
7 & 21.00$\pm$0.03 & -- & 21.30$\pm$0.02 & 21.07$\pm$0.03 & 5$^{+1}_{-1}$ & 14$^{+1}_{-1}$ & 15$^{+1}_{-1}$ \\
8 & 22.80$\pm$0.11 & -- & 21.78$\pm$0.04 & 21.89$\pm$0.05 & 3$^{+1}_{-1}$ & 4$^{+1}_{-1}$ & 4$^{+2}_{-1}$ \\
9 & 20.84$\pm$0.03 & -- & 20.82$\pm$0.04 & 20.74$\pm$0.07 & 5$^{+1}_{-1}$ & 14$^{+1}_{-3}$ & 15$^{+1}_{-2}$ \\
10 & 21.13$\pm$0.04 & -- & 21.43$\pm$0.03 & 21.07$\pm$0.05 & 15$^{+1}_{-1}$ & 14$^{+1}_{-2}$ & 15$^{+1}_{-1}$ \\
11 & 22.89$\pm$0.11 & -- & 22.27$\pm$0.05 & 21.94$\pm$0.13 & 32$^{+94}_{-27}$ & 21$^{+109}_{-15}$ & 20$^{+102}_{-12}$ \\
12 & 23.39$\pm$0.17 & -- & 22.97$\pm$0.09 & 22.71$\pm$0.16 & 35$^{+74}_{-30}$ & 20$^{+79}_{-15}$ & 20$^{+75}_{-13}$ \\
13 & 21.41$\pm$0.05 & -- & 21.51$\pm$0.03 & 21.17$\pm$0.07 & 6$^{+1}_{-1}$ & 14$^{+2}_{-2}$ & 15$^{+1}_{-2}$ \\
14 & 22.30$\pm$0.10 & -- & 21.62$\pm$0.07 & 21.89$\pm$0.22 & 3$^{+3}_{-2}$ & 4$^{+3}_{-2}$ & 5$^{+4}_{-3}$ \\
15 & 21.89$\pm$0.07 & -- & 21.73$\pm$0.04 & 21.04$\pm$0.06 & 15$^{+13}_{-4}$ & 21$^{+3}_{-5}$ & 20$^{+4}_{-4}$ \\
\hline \\ 
\end{tabular} 
\end{table} 
\clearpage
\newpage
\begin{table}
\caption{Cluster $E(B-V)$  and mass estimates. The full table is available in the electronic edition of the journal.} 
\label{tab:clebv}
\begin{tabular}{rllllll} 
\hline \hline 
ID & \ebv & \ebv & \ebv & $M (10^{6}M_{\sun})$ & $M (10^{6}M_{\sun})$ & $M (10^{6}M_{\sun})$\\ 
  &  $Z_{\sun}$ &  $0.4Z_{\sun}$ & $0.2Z_{\sun}$ & $Z_{\sun}$ & $0.4Z_{\sun}$ & $0.2Z_{\sun}$\\ 
\hline
1 & $0^{+0}_{-0}$ & $0^{+0}_{-0}$ & $0^{+0.04}_{-0}$ & 0.029 & 0.029 & 0.029 \\
2 & $0.16^{+0.12}_{-0.16}$ & $0.12^{+0.12}_{-0.12}$ & $0.12^{+0.12}_{-0.12}$ & 0.012 & 0.012 & 0.013 \\
3 & $0.24^{+0.06}_{-0.06}$ & $0.22^{+0.10}_{-0.08}$ & $0.24^{+0.06}_{-0.10}$ & 0.047 & 0.053 & 0.061 \\
4 & $0^{+0.18}_{-0}$ & $0^{+0.06}_{-0}$ & $0^{+0.18}_{-0}$ & 0.013 & 0.013 & 0.013 \\
5 & $0.06^{+0.06}_{-0.04}$ & $0.10^{+0.04}_{-0.04}$ & $0.06^{+0.04}_{-0.04}$ & 0.042 & 0.074 & 0.065 \\
6 & $0^{+0.08}_{-0}$ & $0^{+0.04}_{-0}$ & $0^{+0.04}_{-0}$ & 0.011 & 0.017 & 0.017 \\
7 & $0^{+0}_{-0}$ & $0^{+0}_{-0}$ & $0^{+0}_{-0}$ & 0.081 & 0.074 & 0.051 \\
8 & $0.4^{+0.08}_{-0.08}$ & $0.38^{+0.08}_{-0.08}$ & $0.34^{+0.08}_{-0.08}$ & 0.075 & 0.086 & 0.067 \\
9 & $0^{+0}_{-0}$ & $0^{+0}_{-0}$ & $0^{+0}_{-0}$ & 0.110 & 0.100 & 0.070 \\
10 & $0^{+0}_{-0}$ & $0^{+0}_{-0}$ & $0^{+0}_{-0}$ & 0.081 & 0.074 & 0.081 \\
11 & $0^{+0.24}_{-0}$ & $0.08^{+0.18}_{-0.08}$ & $0.08^{+0.22}_{-0.08}$ & 0.063 & 0.067 & 0.074 \\
12 & $0^{+0.22}_{-0}$ & $0^{+0.22}_{-0}$ & $0^{+0.26}_{-0}$ & 0.024 & 0.024 & 0.039 \\
13 & $0^{+0}_{-0}$ & $0^{+0}_{-0}$ & $0^{+0}_{-0}$ & 0.074 & 0.067 & 0.031 \\
14 & $0.18^{+0.22}_{-0.14}$ & $0.18^{+0.16}_{-0.16}$ & $0.22^{+0.12}_{-0.22}$ & 0.053 & 0.043 & 0.031 \\
15 & $0^{+0.06}_{-0}$ & $0.06^{+0.08}_{-0.06}$ & $0.08^{+0.08}_{-0.08}$ & 0.145 & 0.142 & 0.083 \\
\hline \\ 
\end{tabular} 
\end{table} 
\clearpage
\newpage
\begin{table}
\begin{minipage}{\textwidth} 
\caption{\Hii region photometry} 
\label{tab:hiiphot}
\begin{tabular}{cccccccccccc} 
\hline \hline 
Region & FUV\footnote{\units} & NUV$^a$ & F380W$^a$ & F555W$^a$ & F606W$^a$ & F814W$^a$ & H$\alpha$\footnote{\flux} & $3.6\micron$\footnote{\mJy} & $4.5\micron^c$ & $5.8\micron^c$ & $8.0\micron^c$\\  
\hline 
A & 2080$\pm$95 & 1890$\pm$81 & $549\pm4$ & $396\pm1$ & $459\pm1$ & $276\pm1$ & $139\pm14$ & $2.56\pm0.07$ & $1.58\pm0.04$ & $8.48\pm0.12$ & $24.9\pm0.3$\\
B & 1130$\pm$110 & 474$\pm$63 & $468\pm4$ & -- & $202\pm2$ & $110\pm2$ & $92.9\pm9.4$ & $0.431\pm0.010$ & $0.391\pm0.006$ & $1.01\pm0.05$ & $2.69\pm0.23$\\
C & 1550$\pm$80 & 551$\pm$40 & -- & -- & -- & $153\pm1$ & $52.8\pm5.3$ & $0.235\pm0.019$ & $0.235\pm0.010$ & $0.553\pm0.034$ & $1.20\pm0.10$\\
D & 500$\pm$91 & 236$\pm$58 & $258\pm3$ & -- & $140\pm1$ & $86.2\pm1.7$ & $32.0\pm3.5$ & $0.357\pm0.061$ & $0.255\pm0.048$ & $0.980\pm0.119$ & $2.42\pm0.34$\\
E & $\leqslant218\pm$25 & $\leqslant64.8\pm$21.2 & $15.8\pm3.0$ & -- & $11.2\pm0.7$ & -- & $7.32\pm0.79$ & $\leqslant0.141\pm0.011$ & $\leqslant0.0851\pm0.0150$ & $\leqslant0.365\pm0.050$ & $\leqslant1.37\pm0.09$\\
F & 13600$\pm$100 & 6220$\pm$110 & $13200\pm100$ & $9080\pm10$ & $8610\pm10$ & $5690\pm10$ & $1140\pm110$ & $14.5\pm0.1$ & $12.9\pm0.1$ & $40.2\pm0.2$ & $114\pm1$\\
G & $\leqslant70.3\pm$11.2 & $\leqslant22.4\pm$8.3 & -- & $17.2\pm0.4$ & $16.9\pm0.4$ & $8.32\pm0.28$ & $2.8\pm0.3$ & $0.0152\pm0.0034$ & -- & $0.0552\pm0.0127$ & $0.105\pm0.015$\\
H & $\leqslant93.2\pm$10.4 & $\leqslant43.2\pm$6.5 & -- & $27.6\pm0.4$ & $25.2\pm0.4$ & $10.9\pm0.3$ & $4.87\pm0.51$ & $0.0305\pm0.0035$ & $0.0193\pm0.003$ & $0.0804\pm0.0101$ & $0.154\pm0.011$\\
I & $\leqslant39.2\pm$7.3 & $\leqslant32.0\pm$5.9 & -- & $14.9\pm0.4$ & $14.8\pm0.4$ & $9.73\pm0.27$ & $2.02\pm0.23$ & $0.0259\pm0.0032$ & $0.0186\pm0.003$ & $0.0665\pm0.0107$ & $0.221\pm0.015$\\
J & $\leqslant41.0\pm$6.3 & $\leqslant34.8\pm$5.7 & -- & $10.4\pm0.4$ & $13.0\pm0.4$ & $5.07\pm0.28$ & $3.50\pm0.36$ & $0.0211\pm0.0033$ & $0.0111\pm0.005$ & $0.0609\pm0.0108$ & $0.159\pm0.012$\\
\hline \\ 
\end{tabular} 
\end{minipage}
\end{table}
\begin{table}
\begin{minipage}{\textwidth} 
\label{tab:hiiage}
\caption{\Hii region ages (Myr)}  
\begin{tabular}{ccccccccccccc} 
\hline
\hline 
Region & Age (inst)\footnote{\fnAp} & Age (inst)$^a$ & Age (cont)$^a$ & Age (cont)$^a$ & Med. Age\footnote{\fnMed} & Med Age$^b$ & EW(H$\alpha$)  & EW(H$\alpha$) Age & EW(H$\alpha$) Age & Lit. Age\footnote{\ageref} & $f_{c}$\footnote{\fc} & $L_{H\alpha}/L_{X}$\footnote{\fnLx} \\ 
 & $0.4Z_{\sun}$ & $0.2Z_{\sun}$ & $0.4Z_{\sun}$ &$0.2Z_{\sun}$ & $0.4Z_{\sun}$ &$0.2Z_{\sun}$ & \AA &  $0.4Z_{\sun}$ &$0.2Z_{\sun}$ & & \\ 
\hline
A & $10^{+1}_{-1}$ & $10^{+1}_{-1}$ & $1^{+1}_{-1}$ & $1^{+1}_{-1}$ & $5^{+1}_{-1}$ & $5^{+2}_{-2}$ & $29.8\pm3.3$ & $10^{+0}_{-1}$ & $11^{+0}_{-0}$ & $5\pm0.5$ & $0.18$& 22\\
B & $273^{+65}_{-96}$ & $88^{+16}_{-18}$ & $489^{+273}_{-207}$ & $128^{+1272}_{-42}$ & $14^{+1}_{-3}$ & $15^{+1}_{-2}$ & $39.6\pm5$ & $9^{+0}_{-0}$ & $10^{+0}_{-0}$ & $3.5\pm0.5$ & $0.26$& 11\\
C & --& --& --& --& --& -- & $27.3\pm3$ & $10^{+0}_{-1}$ & $11^{+1}_{-0}$ & $4.5\pm0.5$ & --& 6.1\\
D & $292^{+135}_{-234}$ & $87^{+84}_{-82}$ & $497^{+1103}_{-435}$ & $1200^{+900}_{-1195}$ & $16^{+4}_{-8}$ & $17^{+3}_{-4}$ & $22.4\pm3.4$ & $11^{+1}_{-0}$ & $12^{+1}_{-1}$& -- & $0.17$& $\geqslant23$\\
E & --& --& --& --& --& -- & $56.2\pm21$ & $8^{+2}_{-1}$ & $9^{+1}_{-1}$& -- & --& 3.1\\
F & $222^{+6}_{-6}$ & $86^{+1}_{-1}$ & $421^{+12}_{-11}$ & $2600^{+100}_{-100}$ & $10^{+16}_{-2}$ & $15^{+1}_{-1}$ & $15\pm1.6$ & $12^{+0}_{-0}$ & $14^{+1}_{-1}$ & $5; 15-50$ & $0.25$& 1.5\\
G & $6^{+2}_{-1}$ & $6^{+2}_{-1}$ & $1^{+15}_{-1}$ & $1^{+19}_{-1}$ & $21^{+959}_{-15}$ & $266^{+198}_{-257}$ & $28.5\pm7$ & $10^{+1}_{-1}$ & $11^{+1}_{-1}$& -- & $0.06$& --\\
H & $6^{+1}_{-1}$ & $6^{+2}_{-1}$ & $1^{+8}_{-1}$ & $2^{+8}_{-2}$ & $182^{+36}_{-177}$ & $68^{+24}_{-62}$ & $27.3\pm4.4$ & $10^{+0}_{-1}$ & $11^{+1}_{-0}$& -- & $0.27$& --\\
I & $7^{+376}_{-2}$ & $8^{+1092}_{-2}$ & $1^{+3799}_{-1}$ & $2^{+4298}_{-2}$ & $13^{+12}_{-7}$ & $120^{+74}_{-114}$ & $12.1\pm2.1$ & $13^{+1}_{-1}$ & $16^{+0}_{-1}$& -- & $0.15$& --\\
J & $4^{+1}_{-1}$ & $4^{+1}_{-1}$ & $1^{+7}_{-1}$ & $1^{+8}_{-1}$ & $4^{+1}_{-1}$ & $5^{+1}_{-1}$ & $113\pm82$ & $6^{+3}_{-0}$ & $7^{+3}_{-1}$& -- & $0.31$& --\\
\hline
\end{tabular}
\end{minipage}
\end{table} 
\clearpage
\newpage

\begin{table}
\begin{minipage}{\textwidth} 
\caption{\Hii region extinctions}
\label{tab:hiiebv}
\begin{tabular}{cccccccccc} 
\hline
\hline 
Region & $E(B-V)$ (inst) & $E(B-V)$ (inst) & $E(B-V)$ (cont) & $E(B-V)$ (cont) & Lit. $E(B-V)$\footnote{\ebvref} \\ 
 & $0.4Z_{\sun}$ & $0.2Z_{\sun}$ & $0.4Z_{\sun}$ & $0.2Z_{\sun}$ & & & \\ 
\hline
A & $0.22^{+0.02}_{-0.02}$& $0.14^{+0.02}_{-0.02}$& $0.36^{+0.02}_{-0.02}$& $0.34^{+0.02}_{-0.02}$ & $0.25$ \\
B & $0^{+0.06}_{-0}$& $0.12^{+0.08}_{-0.08}$& $0.12^{+0.08}_{-0.10}$& $0.18^{+0.08}_{-0.18}$ & $0.28$ \\
C & --& --& --& -- & $0.34$ \\
D & $0.10^{+0.14}_{-0.10}$& $0.20^{+0.08}_{-0.20}$& $0.18^{+0.12}_{-0.18}$& $0^{+0.48}_{-0}$& -- \\
E & --& --& --& --& -- \\
F & $0.18^{+0.02}_{-0.18}$& $0.28^{+0.02}_{-0.02}$& $0.26^{+0.02}_{-0.02}$& $0^{+0}_{-0}$ & $0.22; 0.08$ \\
G & $0.24^{+0.08}_{-0.24}$& $0.42^{+0.08}_{-0.20}$& $0.22^{+0.14}_{-0.08}$& $0.20^{+0.16}_{-0.08}$& -- \\
H & $0.12^{+0.06}_{-0.06}$& $0.28^{+0.06}_{-0.18}$& $0.08^{+0.14}_{-0.06}$& $0.08^{+0.12}_{-0.06}$& -- \\
I & $0.24^{+0.30}_{-0.16}$& $0.44^{+0.14}_{-0.44}$& $0.44^{+0.16}_{-0.44}$& $0.44^{+0.44}_{-0.44}$& -- \\
J & $0.42^{+0.12}_{-0.12}$& $0.38^{+0.12}_{-0.12}$& $0.14^{+0.16}_{-0.12}$& $0.12^{+0.16}_{-0.12}$& -- \\
\hline
\end{tabular}
\end{minipage}
\end{table}

\begin{table}
\begin{minipage}{\textwidth} 
\caption{Clusters by age bin for $Z = 0.2Z_{\sun}$ } 
\label{tab:bin02}
\begin{tabular}{rccccc} 
\hline \hline 
Age & All & Luminosity Cut & Mass Cut & CCF$_{30}$\footnote{\ccfthirty} & CCF$_{100}$\footnote{\ccfhundred}\\ 
 
\hline 
1--10 Myr & 59 & 23 & 13 & 10 & 10\\
11--22 Myr & 70 & 28 & 28 & 12 & 12\\
23--100 Myr & 2 & 0 & 1 & 8 & 78\\
$> 100$ Myr & 27 & 2 & 21 & 0 & 0\\
\hline
Total & 158 & 53 & 63 & 30 & 100\\

\hline \\ 
\end{tabular}
\end{minipage}
\end{table}


\label{lastpage}

\end{document}